\documentclass{aastex61}
\pdfoutput=1 
\usepackage{amsmath,amstext}
\usepackage[T1]{fontenc}
\usepackage[figure,figure*]{hypcap}

\usepackage{graphicx}
\usepackage{epstopdf}
\usepackage{rotating}
\usepackage{latexsym}
\usepackage{amssymb}
\usepackage{longtable} 
\usepackage{epsf}
\usepackage{natbib}

\received{}
\revised{}
\accepted{}

\submitjournal{The Astrophysical Journal}

\shorttitle{Stellar Population in the High-Mass Star-Forming Region IRAS$\,$16562-3959}
\shortauthors{Montes et al.}

\begin{document}
\title{A Chandra X-ray and Infrared Study of the Stellar Population in the High-Mass Star Forming Region IRAS 16562-3959}

\author{Virginie A. Montes}
\affil{Physics Department, New Mexico Institute of Mining and Technology, 801 Leroy Place, Socorro, NM 87801, USA}

\author{Peter Hofner}
\affiliation{Physics Department, New Mexico Institute of Mining and Technology, 801 Leroy Place, Socorro, NM 87801, USA}
\affiliation{Adjunct Astronomer at the National Radio Astronomy Observatory}

\author{Lidia M. Oskinova}
\affiliation{Institute for Physics and Astronomy, University of Potsdam, D-14476 Potsdam, Germany}

\author{Hendrik Linz}
\affiliation{Max-Planck-Institute for Astronomy, K\"{o}nigstuhl 17, 69117 Heidelberg, Germany}

\correspondingauthor{Virginie A. Montes}
\email{virginie.montes@student.nmt.edu}

\begin{abstract}
We present the results from Chandra X-ray observations, and near- and mid-infrared analysis, using VISTA/VVV and Spitzer/GLIMPSE catalogs, of the high-mass star-forming region IRAS\,16562-3959, which contains a candidate for a
high mass protostar. We detected 249 X-ray sources within the ACIS-I field-of-view. The majority of the X-ray sources have low count rates ($<$ 0.638\,cts/ks), and hard X-ray spectra. The search for YSOs in the region using VISTA/VVV and Spitzer/GLIMPSE catalogs resulted in a total of 799 YSOs, with 66 Class I and 733 Class II YSOs. The search for near- and mid-infrared counterparts of the X-ray sources led to a total of 165 VISTA/VVV counterparts, and a total of 151 Spitzer/GLIMPSE counterparts. The infrared analysis of the X-ray counterparts allowed us to add an extra 123 YSOs associated with the region. We conclude that a total of 922 YSOs are associated with the region, with 67 Class I, 764 Class II, and 91 Class III YSOs. We also found that the region is composed of 11 subclusters. In the vicinity of the high-mass protostar, the stellar distribution has a core-halo structure. The subcluster containing the high-mass protostar is the densest and the youngest in the region, and the high-mass protostar
is located at its center. The YSOs in this cluster appear to be substantially older than the high mass protostar.

\end{abstract}


\section{Introduction}
\label{sec:Introduction}

High-mass stars play a dominant role in the physical, chemical, and morphological structure of their host galaxies \citep[e.g.][]{Kennicutt98, Cesaroni05}, but despite their importance, the formation and early evolution of high-mass stars is still not well understood. The two main theories trying to explain high-mass star formation are the monolithic core accretion model \citep{McKee03} and the competitive accretion model \citep{Bonnell01}. In the monolithic core accretion model,
the physical properties of the molecular core determines the final mass of the star, and it will form in isolation, or in small-N multiple systems. In contrast, in the competitive accretion model, the high-mass star forms at the center of a low-mass stellar cluster, and the mass of the high-mass star is related to the properties of the cluster. See the recent
review of \citet{Tan14} for more details. Characterizing the stellar population associated with high-mass star formation is crucial to decide between the two main theories. Studies of more evolved stellar systems indicate that most high-mass stars are in fact formed in a dense cluster environment \citep{Lada03}, but there is a small population of high mass stars
which appear to have formed in isolation \citep{dewit05, Oskinova13}).
As dynamical evolution can significantly change the appearance of a young stellar system, studying early stages of high-mass star formation in young, and hence embedded regions is critical. 

Many previous studies of high-mass star forming regions have been carried out at infrared wavelengths \cite[e.g.][]{Lada03} because long wavelengths can penetrate the dust efficiently. However, observing only at infrared wavelengths prevents the detection of YSOs which have partially, or completely lost their circumstellar disk. The Chandra X-ray Observatory allows an alternative view of young stellar populations, as X-ray extinction is low at energies above a few keV, and the background/foreground contamination is relatively small, mainly from nearby stars and AGNs.
Furthermore, Classical T Tauri Stars (CTTS) and Weak-line T Tauri Stars (WTTS) are bright in X-rays \citep{Feigelson99}, and earlier evolutionary phases of YSOs (Class I) are also well known X-ray emitters \citep{Grosso97, Neuhauser97}. Thus, combining X-ray data with infrared data is an efficient way to study the stellar population in regions where massive stars are forming. In this paper, we present such a study for the high-mass star forming region IRAS\,16562-3959, where we have combined Chandra X-ray images and spectra with near- and mid-infrared data from the
VISTA/VVV survey, and the Spitzer GLIMPSE survey. 

The high-mass protostellar candidate G345.4938+01.4677 associated with IRAS 16562-3959 is located at a distance of $1.7$ kpc, and the region has a FIR luminosity of L = $7\times 10^4$\,L$_\odot$, clearly indicating the presence of
high mass (proto) stars \citep{Lopez11}. It was discovered by \citet{Guzman10} at $6\,$cm with ATCA, while searching for ionized jets in regions of high FIR luminosity. The $6\,$cm ATCA data show five aligned continuum sources, where the bright central source presumably indicates the position of the high-mass protostar. The four symmetrically displaced sources have been observed to move away from the central source at high speed \citep{Guzman14}, and are likely due to shock-ionization from fast outflowing matter in a jet. IRAS$\,$16562-3959 is also associated with a slower ionized wind, an infalling envelope, and a bipolar molecular outflow, and the mass of the central object has been estimated as $\sim15$ M$_\odot$ \citep{Guzman14}. Recently, \citet{Lopez16} reported $^{13}$CO(3-2) APEX observations of this region, which showed that the high-mass protostellar candidate is located at the column density maximum, further demonstrating its youth.

It is important to point out that while many similar studies have been
performed in high-mass star forming regions (e.g. M17, W3, see e.g. \citet{Hofner02, Townsley14}) where H{\small II} regions clearly indicate the presence of fully formed massive stars, the IRAS\,16562-3959 region contains only a weak radio source which is clearly in a pre-H{\small II} region phase of evolution, and is also the only manifestation of high mass star formation within several pc. It thus allows to study the stellar population associated with the formation of a high-mass star in a very early evolutionary phase.

In Section~\ref{sec:Chandra X-Ray Observations and Data Reduction}, we describe the Chandra observations and data reduction of the region. In Section~\ref{sec:X-Ray Results and Analysis}, the X-ray results and analysis of the region are described. A near- and mid-infrared analysis of the region is presented in Section~\ref{sec:Infrared Analysis of the Region}.  X-ray and infrared source associations are discussed in Section~\ref{sec:X-ray and Infrared Associations}. In Section~\ref{sec:Subcluster Analysis}, we present a subcluster analysis of the large scale region. In Section~\ref{sec:Discussion}, we give a discussion of the observational results, and in Section~\ref{sec:Summary} a summary, and general conclusions.

\section{Chandra X-Ray Observations and Data Reduction}
\label{sec:Chandra X-Ray Observations and Data Reduction}

The IRAS\,16562-3959 high-mass star-forming region was observed with the Chandra X-Ray Observatory at three different epochs detailed in Table~\ref{tab:log_chandra_obs}. The first observation (Obs$\#\,1$) was made in cycle $14$ with an exposure time of 5.02 ks, the second (Obs$\#\,2$) and third observations (Obs$\#\,3$) were obtained in cycle 16 with exposure times of $38.58$\,ks and $40.07$\,ks respectively, corresponding to a total exposure time of $83.67$\,ks, i.e. approximately $23$\,hr. No background flares were detected during these three
epochs. The observations were carried out using the imaging array of the Advanced CCD Imaging Spectrometer camera (ACIS-I). ACIS-I consists of four 1024 x 1024 pixel CCDs covering a field-of-view of $17\arcmin$ x $17\arcmin$, and has an energy range from $0.1$ to $10$\,keV. For details on the instrument see \citet{Weisskopf96, Weisskopf02, Garmire03}. The nominal pointing position used for the three observations was R.A.(J2000) = $16^h59^m41\fs60$, Dec.(J2000) = $-40^\circ03\arcmin43\farcs6$, which are the coordinates of the high-mass protostar \citep{Guzman10}. The satellite roll angle (i.e., the angle between the celestial north an the Z-axis of the spacecraft) for Obs$\#\,1$ was $95.45^\circ$ and $300.22^\circ$ for Obs$\#\,2$ and $3$.

  Data reduction was performed using the CIAO software package version 4.7 and CALDB 4.6.8 provided by the Chandra X-ray Center. The data were recalibrated using the \textit{chandra$\_$repro} reprocessing script to create a new level=2 event file and a new bad pixel file for each observation. 
  Before any analysis, it is necessary to correct the absolute astrometry on the three observations. For this, we first used the Obs$\#\,2$ data to match Chandra sources with NIR sources from the 2MASS catalog. A set of $23$ bright sources were selected in the center of the field-of-view of the ACIS-I array and compared to their 2MASS counterparts. A systematic offset of $0\farcs1$ towards South and $0\farcs35$ toward West was observed. After astrometry correction, the rms offset between Chandra Obs$\#\,2$ and 2MASS sources was about $0\farcs1$.
Subsequently we corrected the astrometry of Obs$\#\,1$ and $3$ to match the position of the 23 X-ray sources (when detected) of Obs$\#\,2$.

Finally, we filtered the data in three different energy ranges: $0.5 - 2$\,keV (soft band), $2 - 8$\,keV (hard band) and $0.5 - 8$\,keV (full band) and created flux images, exposure maps and PSF maps for each observation and each energy range.

\section{X-Ray Results and Analysis}
\label{sec:X-Ray Results and Analysis}

\subsection{Detection}
The source identification in the ACIS-I field-of-view was performed using \textit{wavdetect}, a wavelet-based source detection algorithm \citep{Freeman02}, in three energy ranges: $0.5 - 2$\,keV (soft band), $2 - 8$\,keV (hard band) and $0.5 - 8$\,keV (full band). We used a threshold significance of $10^{-6}$ corresponding to one spurious source in the field-of-view of ACIS-I, and wavelet scale sizes from $1$ to $16$ pixels incremented by a factor $\sqrt{2}$. The search was done separately in the three observations and the source lists were joined. Sources with at least $5$ counts in the combined observations were considered to be real sources. Due to the large variation of the extinction in the region (our spectral fits imply a range of N$_H$ between $2\times 10^{21} - 5\times 10^{23}\,$cm$^{-2}$; see below), this detection limit corresponds to a range of limiting luminosities of
L$_{X,lim} = 2.5 \times 10^{29} - 1.6 \times 10^{32}$\,erg.s$^{-1}$, for a thermal plasma with $kT = 1$\,keV.

\begin{figure}
\plotone{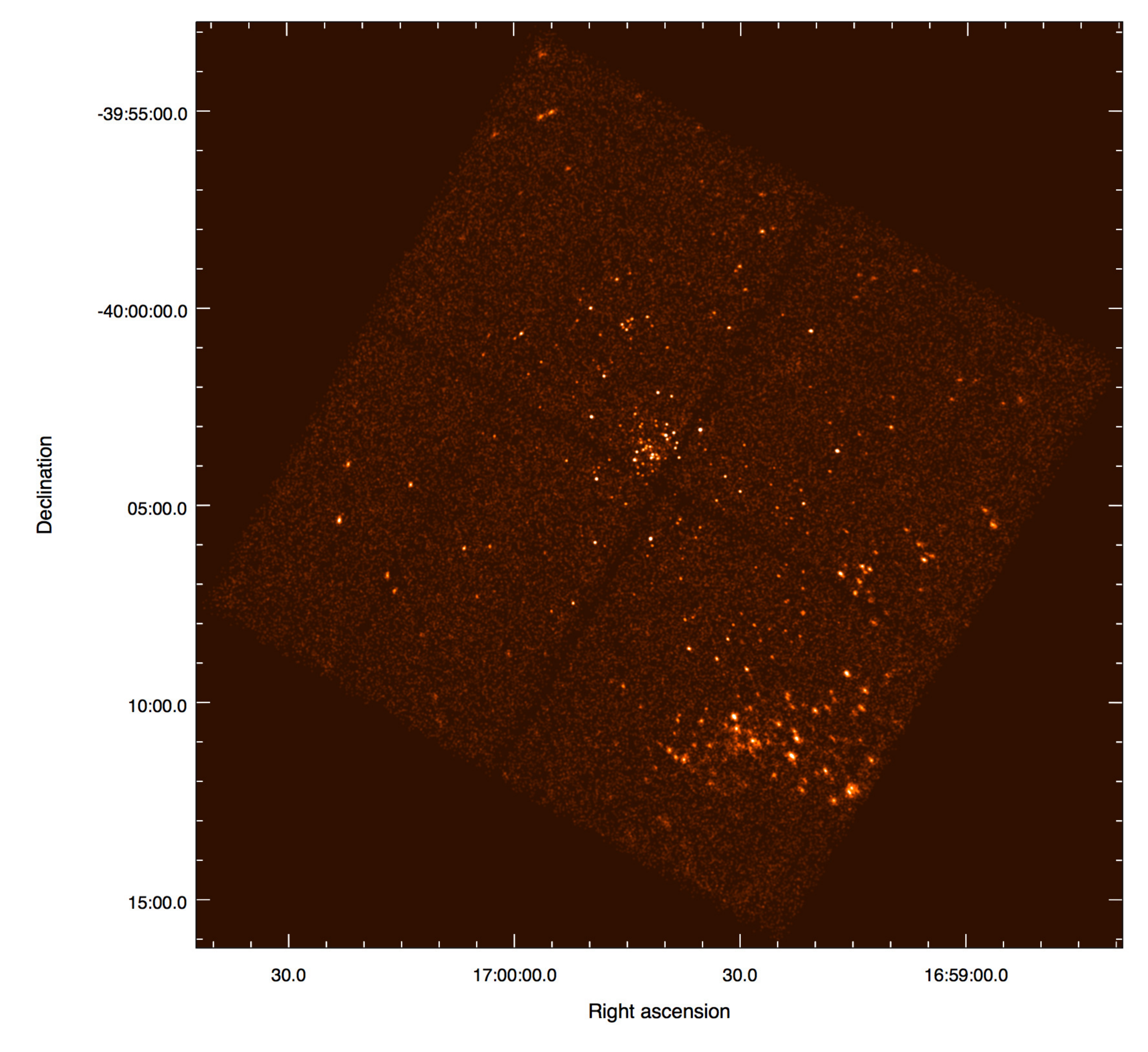}
\caption{Color image of the three combined observations for the full $17^\prime \times 17^\prime$ ACIS-I field in the $0.5 - 8\,$keV band toward IRAS$\,16562-3959$. A total of 249 X-ray point sources were detected.}
\label{fig:chandra}
\end{figure}

A total of 249 sources were detected in the field-of-view for the full energy range in the combined three observations, as shown on the flux image in Figure~\ref{fig:chandra}. In Obs$\#\,1$, a total of $38$ sources were detected in the full band, we found $19$ sources in the soft band and $20$ sources in the hard band. In Obs$\#\,2$, a total of $208$ sources were detected in the full range of energy, with $102$ sources for the soft band and $147$ sources for the hard band. In the Obs$\#\,3$, a total of $198$ sources were detected in the full band, $106$ sources were detected in the soft band and $148$ sources in the hard band. 
In Table~\ref{tab:X-ray_sources} we list the observed X-ray properties of all our detections. Column~(1) lists the source number, column~(2) and (3) give the Chandra positions, column~(4) is the observed count rate, and column~(5) shows the observed X-ray flux. In column~(6) we list the hardness ratio for each source defined as $HR=\frac{h_{x}-s_{x}}{h_{x}+s_{x}}$ where $h_{x}$ is the count rate in the $2 - 8$\,keV energy range, and $s_{x}$ is the count rate in the $0.5 - 2$\,keV energy range. In column~(7) we give the short term variability for each source in each observation, column~(8) shows the long term variability of each source, and column~(9) lists for each source whether it is variable and which type of variability it shows. Source variability will be discussed in Section~\ref{subsec:Timing Analysis and Source Variability}. As expected by studies of stellar X-ray emission in highly embedded clusters at kpc distances, most
sources are fairly faint, with approximatively $70\,\%$ of the sources having less than $50$ counts (count rate $0.638$\,cts/ks), and according to our hardness ratio classification, the majority of sources have a relatively hard observed X-ray spectrum.

\subsection{Timing Analysis and Source Variability}
\label{subsec:Timing Analysis and Source Variability}
As we have observed this star-forming region at three different epochs, we want to determine if the sources in the field exhibit short-term and/or long-term variabilities. We will refer to changes in count rate of a source within one observation to short term variability. On the other hand, long-term variability corresponds to the overall change in count rate of a source between observations. A number of physical mechanisms which could explain the different types of variability have been
discussed in \citet{Flaccomio12}. 

To determine if the sources display short term variability, we first apply the barycenter correction on each observation. Then we compute the fractional area of each source, which creates a correction for instrumental effects, needed for input to the \textit{glvary} tool. Subsequently we ran \textit{glvary} for each individual source, on each observation subdivided into temporal bins of approximate width $2\,$ks. The \textit{glvary} script uses the Gregory-Loredo variability test algorithm \citep{GregoryLoredo92} which is insensitive to the lightcurve shape and does not overinterpret data in low count rate source as the Kolmogorov-Smirnov (K-S) test could do\footnote{CIAO manual}. This test gives us a variability index indicating if the source is variable or not as shown in Table~\ref{tab:X-ray_sources} column~7. If the variability index (S.T. Var) is between $0$ to $3$, it is unlikely that the source is variable.
If S.T. Var is $4$ or $5$, the source may be variable, and if S.T. Var is above $5$, the source is definitely variable \citep{Rots06}.

\begin{figure}
\plotone{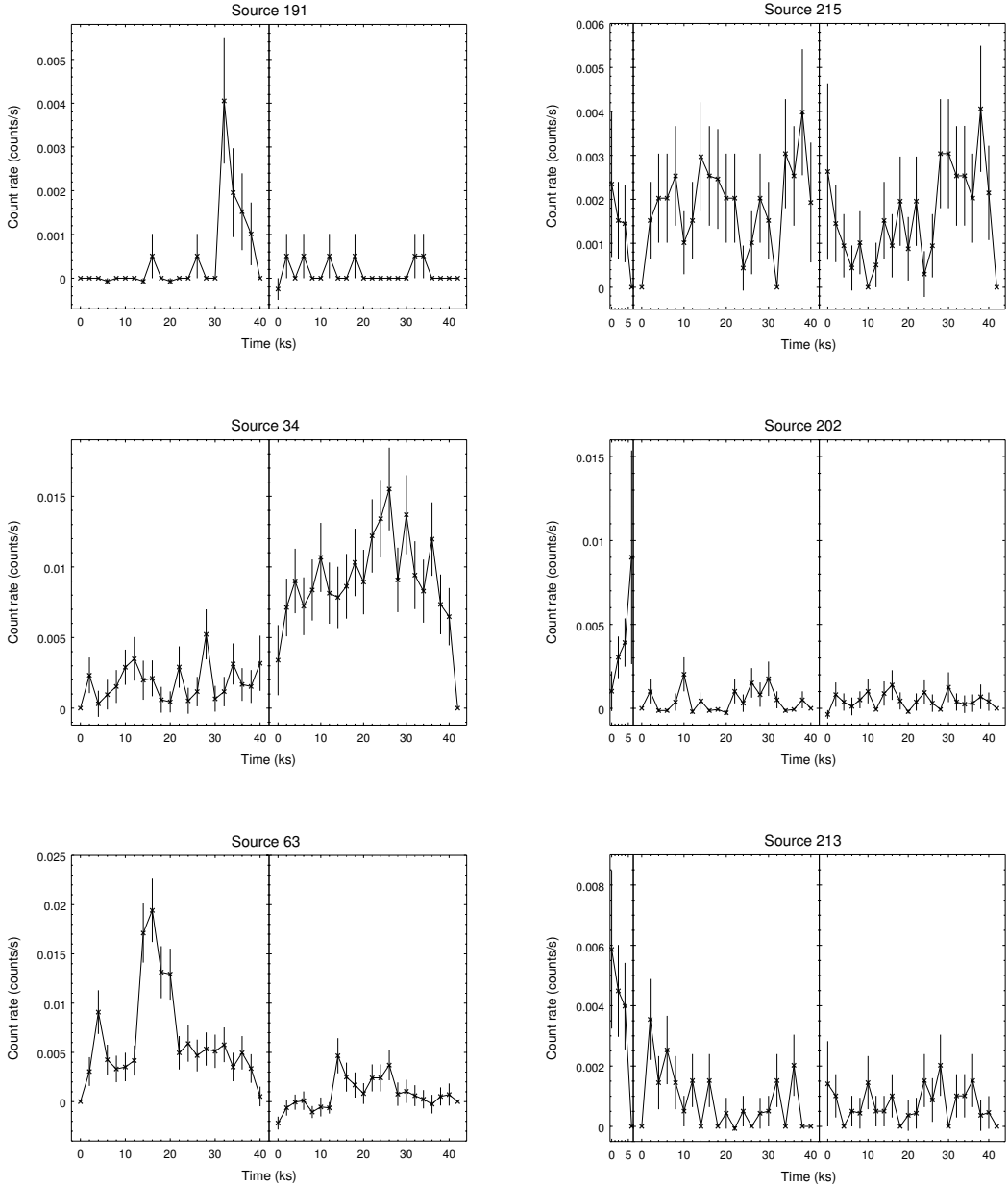}
\caption{Sample light curves of sources exhibiting short-term and/or long-term variability. The three observations are plotted next to each other.
Note that sources on the left panel were not detected in the Obs $\#$1, hence we are only showing the light curves corresponding to Obs $\#$2 and 3. Sources $\#$191 and 215 exhibit short-term variability. For source $\#$191, a flare is observed at the end of Obs $\#$2. Sources $\#$34 and 202 exhibit long-term variability, with markedly increased X-ray activity during Obs $\#$3 for source $\#$34, and in Obs $\#$1 for source $\#$202. Sources $\#$63 and 213 exhibit both short-term and long-term variability. Source $\#$63 shows possible flares in both observations and a higher base X-ray activity in Obs  $\#$2. Source $\#$213 shows a short-term variability during Obs $\#$2 and a higher X-ray emission level in Obs  $\#$1.}
\label{fig:lightcurves}
\end{figure}

To determine if the sources have long-term variability, we follow the method from \citet{Fridriksson08} by computing a significance parameter S (L. T. Var. parameter) for long-term flux variability  for each source:

\begin{equation}
S = \frac{|F_{max}-F_{min}|}{\sqrt{\sigma^2_{F_{max}}+\sigma^2_{F_{min}}}}
     \end{equation}
     
where F$_{max}$ and F$_{min}$ are the maximum and minimum X-ray count rates during the three observations for each source, and $\sigma^2_{F_{max}}$ and $\sigma^2_{F_{min}}$ are the corresponding errors. A source is defined as long-term variable if S $> 3$. The long-term variability parameter for each source is listed in Table~\ref{tab:X-ray_sources} column~(8). In Figure~\ref{fig:lightcurves}, we show a sample of sources exhibiting short-term and/or long-term variability.

\subsection{Spectroscopy}
\label{subsec:spectroscopy}
\begin{figure}
\centering
\includegraphics[scale=0.18]{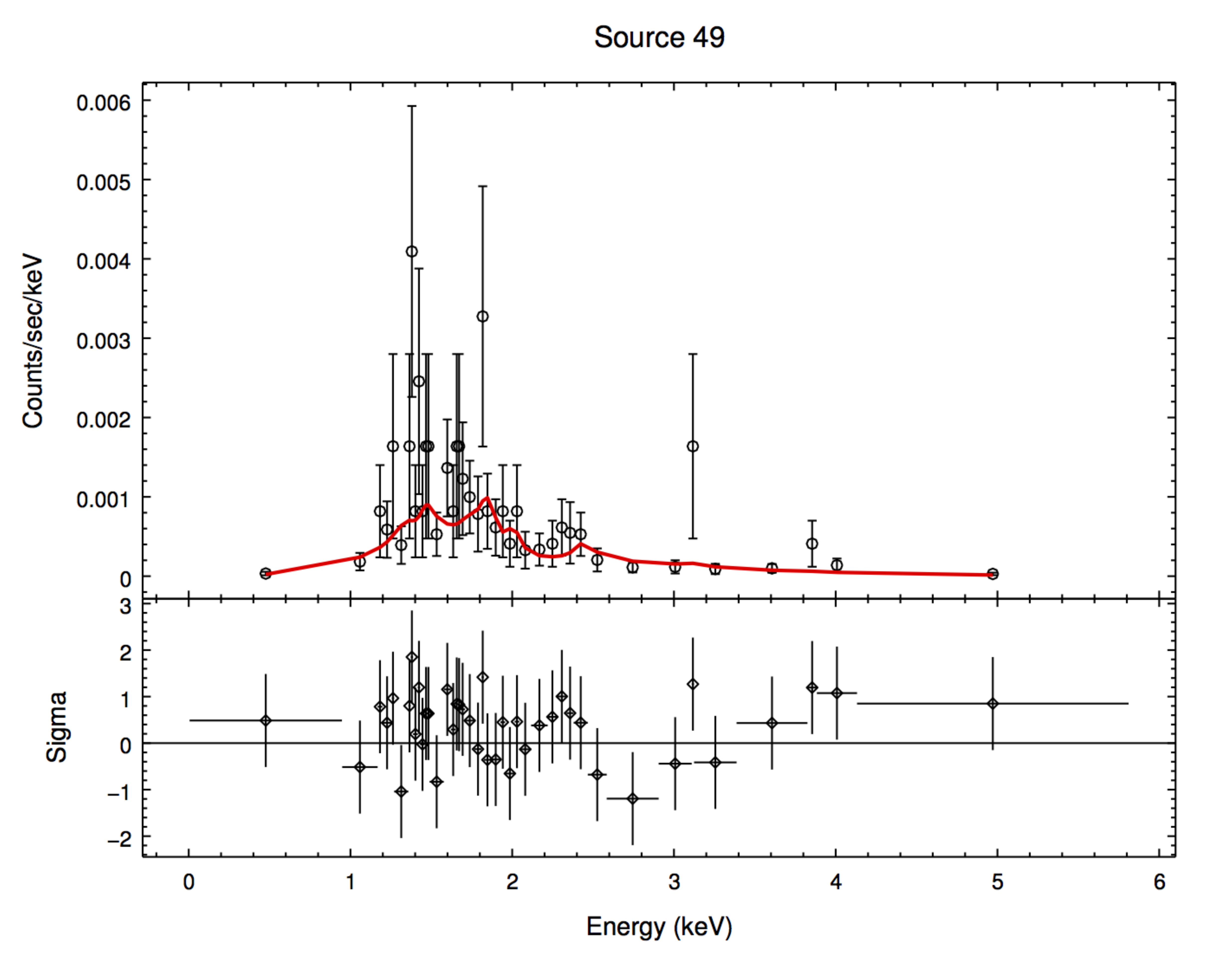}\includegraphics[scale=0.18]{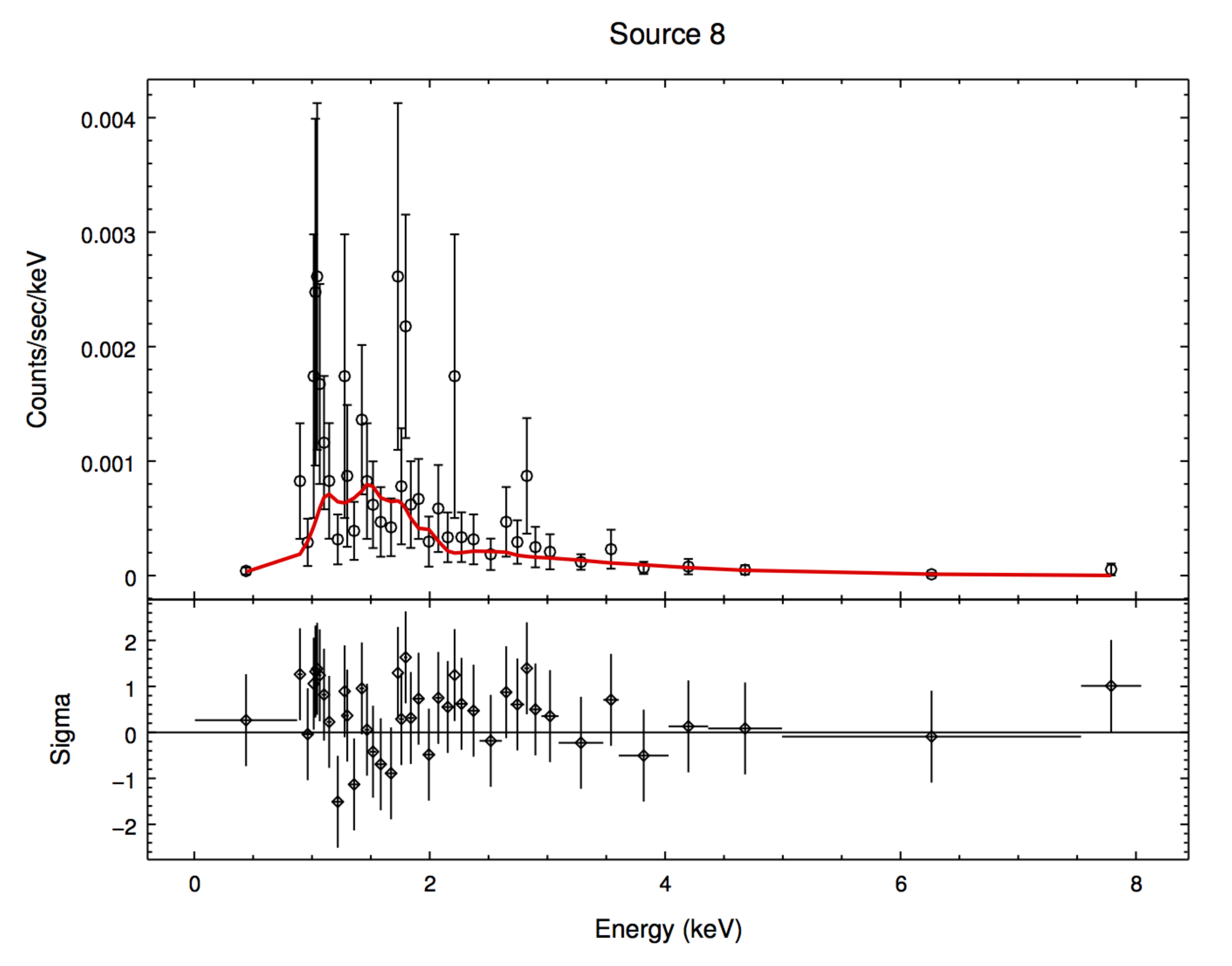}
\caption{Sample spectra as described in the text. For source $\#$49 we were able to obtain a simultaneous fit of N$_H$, T, and Emission Measure, whereas for
Source $\#$8 the temperature was frozen in the fit.}
\label{fig:spectra}
\end{figure}

A spectrum for each source and for each observation was extracted with \textit{srcflux}. To get a single spectrum for each source for all observations, the coadding of the spectra 
and response files was done using \textit{combine$\_$spectra}. The fitting of each coadded spectrum was then run using Sherpa v1. We used an Absorption $\times$ 1-Temperature 
model using the $xswabs$ and $xsapec$ models, with an abundance frozen to 0.3 solar abundance in all cases, which is typical for YSOs \citep{Imanishi01}. Due to the weakness of the sources, we were only able to obtain meaningful fit 
results for 98 sources, using the following method: after a first initial simultaneous fitting of N$_H$, T and Emission measure (EM), we kept 
the fit results if the uncertainties for the parameters were reasonable. This was the case for $6$ sources only.
If the best model fit showed large uncertainty for the fitted parameters, we froze the fitted value for T and
redid the fit. This was the case for $88$ sources. In $4$ cases the value for the temperature resulted initially in unreasonable high values, and in these cases we
assumed and froze a value for T, and fitted for the other parameters. All the fitting results are displayed in Table~\ref{tab:X-ray_properties}, and sample spectra are shown in Figure~\ref{fig:spectra}.

\section{Infrared Analysis of the Region}
\label{sec:Infrared Analysis of the Region}
In order to carry out a more complete analysis of the evolution of the YSO population in the region, we added to our set of X-ray data, all infrared sources detected within a
distance of $15\arcmin$ from the high-mass protostar, using mid-infrared ([3.6], [4.5], [5.8] and [8.0]$\,\micron$) data from the Spitzer/GLIMPSE\footnote{\url{http://irsa.ipac.caltech.edu/}} \citep{Benjamin03} and
near-infrared (J, H, and K$_s$ bands) VISTA/VVV\footnote{\url{http://horus.roe.ac.uk/vsa/}} \citep{Minniti10} surveys. A total of $17422$ mid-infrared sources were found in the Spitzer/GLIMPSE catalog, and $136183$ near-infrared 
sources were found in the VISTA/VVV catalog. This set of data allows in principle to identify YSOs with disks (Class I and Class II).

\begin{figure}
\plotone{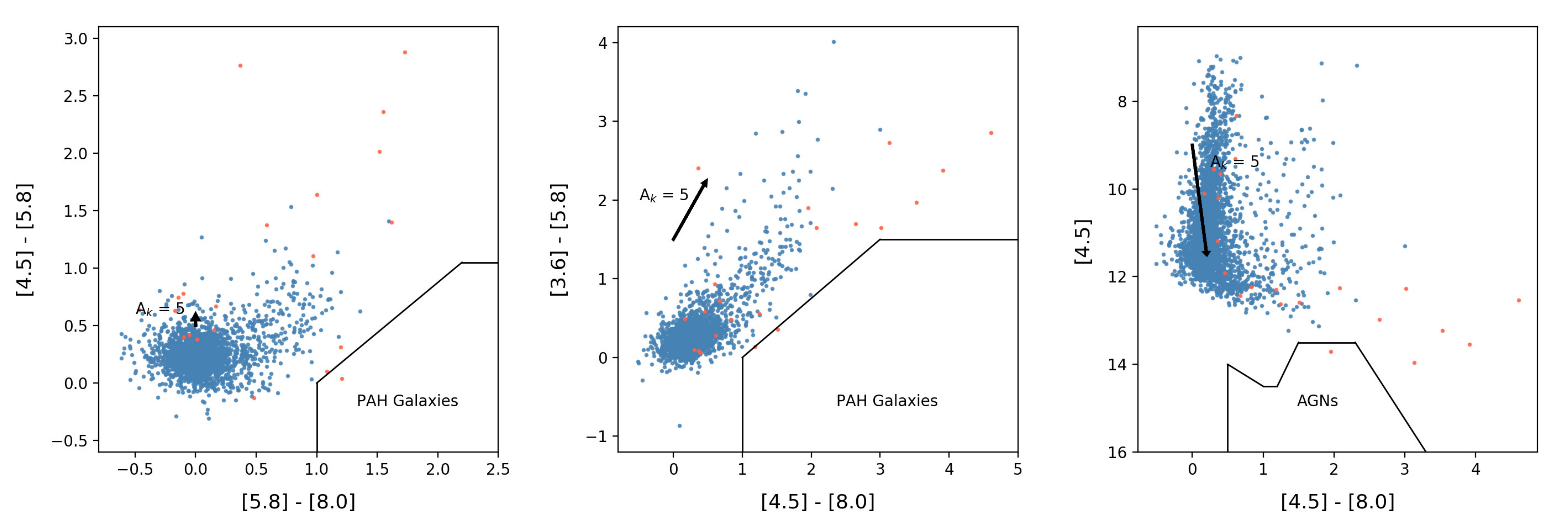}
\caption{Color-color diagrams (CCDs) used for the determination of the contamination due to star-forming galaxies (left and middle panels); and AGNs (right panel), see \citet{Gutermuth09} for more details. Orange dots correspond to all contamination found in Phase$\,$I; blue dots are field stars.}
\label{fig:phaseI_gutermuth}
\end{figure}

\begin{figure}
\plotone{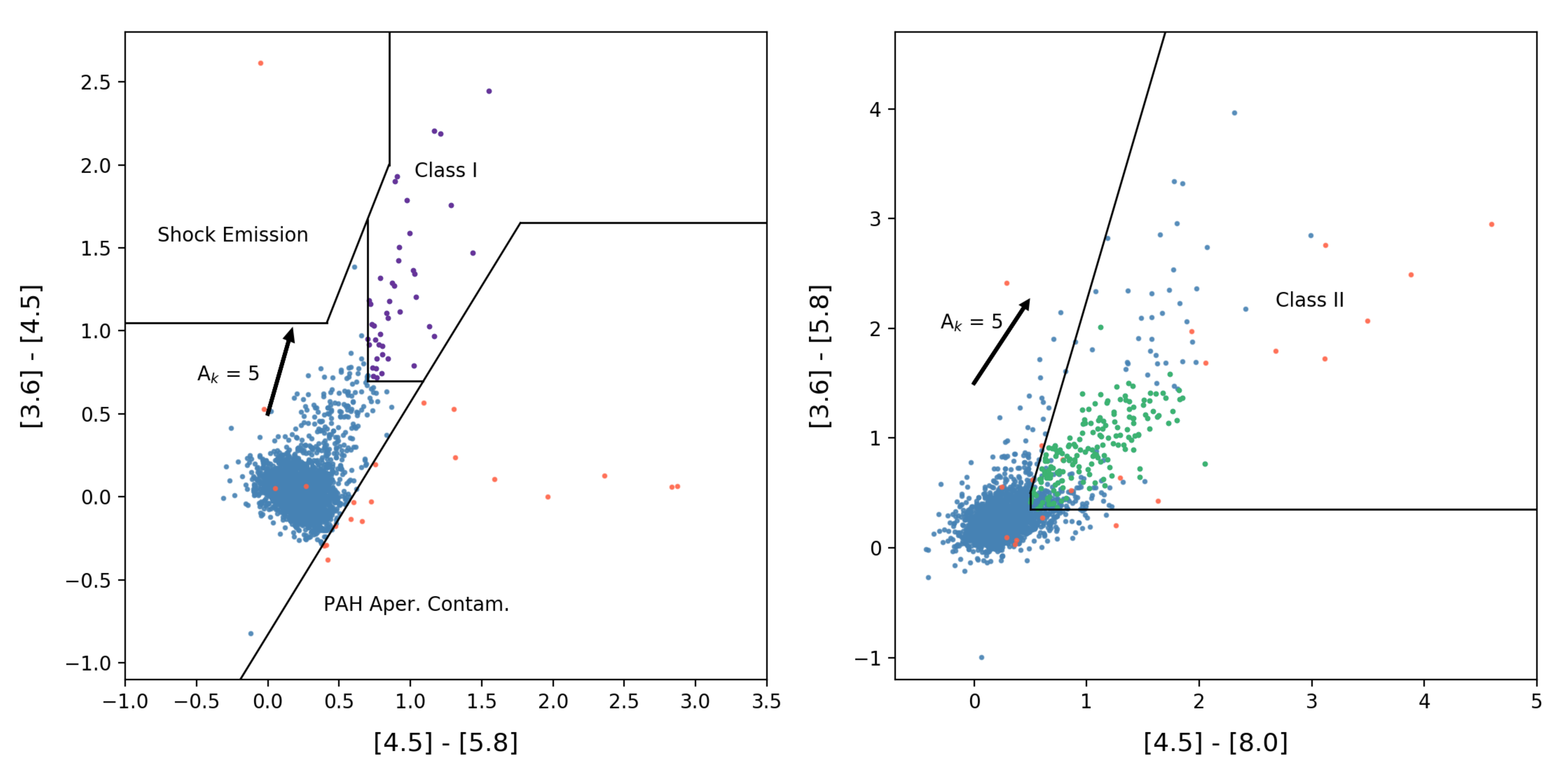}
\caption{CCDs used to determine contamination and classification of sources. The left panel shows the criteria to isolate unresolved shock emission knots
and sources with PAH aperture contamination, and also defines Class$\,$I protostars (purple dots). The right panel shows the location of Class$\,$II YSOs (green dots).
Orange dots correspond to all contamination found in Phase$\,$I, blue dots are field stars.}
\label{fig:phaseI_classification}
\end{figure}

\begin{figure}
\plotone{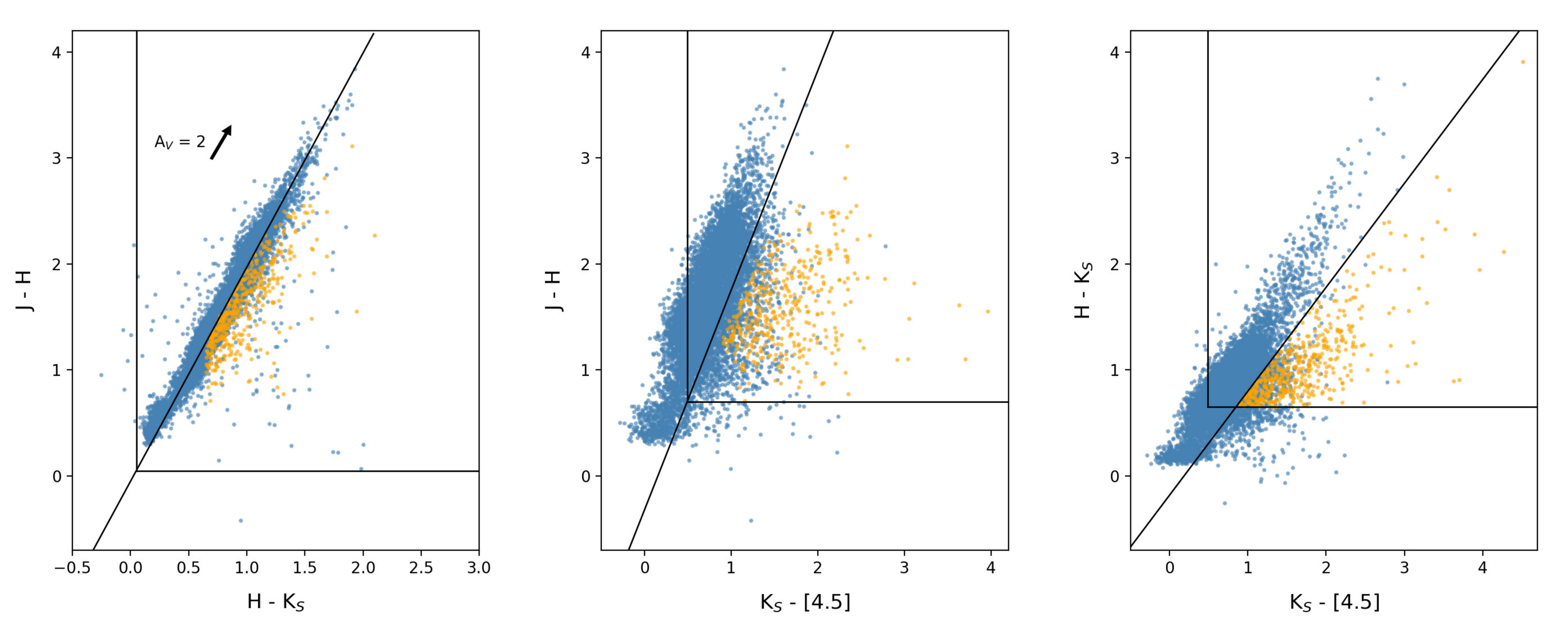}
\caption{CCDs showing the infrared excess sources selection. The orange dots correspond to the infrared excess sources, and the blue dots correspond to field stars. \textit{Left:} Color-color diagram J-H vs H-K$_S$. The black lines mark the limit for the infrared excess selection with (J-H) < (H-K$_S$) $\times$ 2.02 - 0.299 \citep{Soto13}, (J-H) > 0.05 mag, and (H-K$_S$) > 0.05 mag \citep{Zeilder16}. \textit{Middle:} CCD J-H vs K$_S$-[4.5]. The black lines mark the limit for the infrared selection with (J-H) < [(K$_S$-[4.5])-0.49] $\times$ 2.07 + 0.7, (J-H) > 0.7, and (K$_S$-[4.5]) > 0.49 \citep{Zeilder16}. \textit{Right:} CCD H-K$_S$ vs K$_S$-[4.5]. The black line mark the limit for the infrared selection with (H-K$_S$) < [(K$_S$-[4.5])-0.49] $\times$ 0.98 + 0.3, (H-K$_S$) > 0.65, and (K$_S$-[4.5]) > 0.49 \citep{Jose16}.}
\label{fig:phaseII_gutermuth}
\end{figure}

\begin{figure}
\plotone{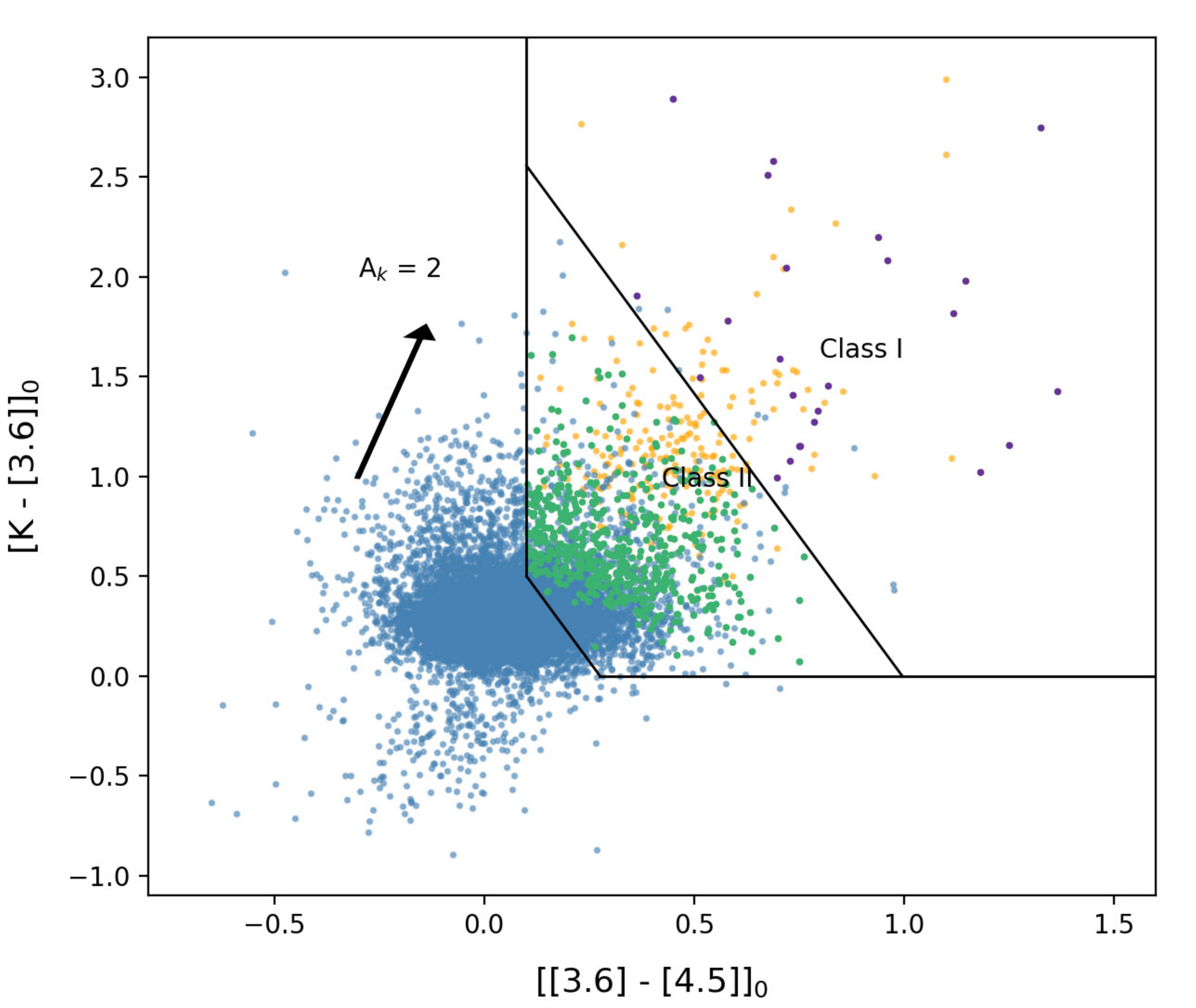}
\caption{CCD [K$_S$-[3.6]]$_0$ vs [[3.6]-[4.5]]$_0$ used for the isolation of Class I and Class II YSOs. Purple dots correspond to Class I YSOs, green dots correspond to Class II YSOs, yellow dots are infrared excess sources, and blue dots are field stars.}
\label{fig:phaseII_classification}
\end{figure}

To determine foreground and background contamination by star forming galaxies, AGNs, shock emission, and extended PAH emission, as well as which stars are
field stars, and which are Class I or II YSOs, we performed the infrared color selection method described in \citet{Gutermuth09}. In Phase$\,$I of this method, we used only GLIMPSE sources that have photometry in all four IRAC bands, and have photometric 
uncertainties $\sigma$ < 0.2 mag in all four bands, which corresponds to a total of $2723$ sources. In Figure~\ref{fig:phaseI_gutermuth} we show Color-Color Diagrams (hereafter CCDs) from the first step of the Phase$\,$I selection method, that allowed us to identify contamination from star-forming galaxies and AGNs.
Then, we proceeded to the elimination of shock emission and extended PAH emission contamination, as well as YSO class selection as shown on Figure~\ref{fig:phaseI_classification}.
We ended with a total of 2702 Spitzer/GLIMPSE sources without contamination. We obtained a total of 42 sources classified as Class I, and 176 sources as Class II.
We continued with a process similar to the Phase$\,$II selection method described by \citet{Gutermuth09}, which we slightly modified due to the use of 
VISTA/VVV data instead of 2MASS data used in the \citet{Gutermuth09} paper. In this step, both Spitzer/GLIMPSE and VISTA/VVV data were used. 
Specifically, the Phase$\,$II selection method is applied to Spitzer/GLIMPSE sources that lack [5.8] and/or [8.0] detections. The Spitzer/GLIMPSE sources were first matched with their VISTA/VVV counterparts. We only selected high quality VISTA/VVV detections with $\sigma<$ 0.1 mag, and Spitzer/GLIMPSE detections with photometric uncertainties $\sigma<$ 0.2 mag in the detected bands for this analysis.
A cross-match of both catalogs was done using a search radius of 1$\arcsec$, to create a matched list of 13309 sources.
A further selection on these sources was done using their VISTA/VVV magnitudes. We excluded stars with saturated photometry \citep{Soto13}, by limiting the magnitudes of the detected VISTA/VVV sources to 13.8, 12.8, and 12.8, for J-, H- and K$_S$-band, respectively. A total of 9974 non-saturated sources were selected. Whenever possible we checked for contamination using similar color-color criteria as described for Phase$\,$I, and ended up with a total of 9844 sources without contamination.
From these sources, we selected stars exhibiting infrared excess by three different color selection diagrams, as shown in Figure~\ref{fig:phaseII_gutermuth}.
The first selection was done using the CCD J-H vs H-K$_S$, where we adopted for the slope of the reddening band $\frac{E_{J-H}}{E_{H-K_{S}}}= 2.02 \pm 0.04$ in the original VISTA 
system \citep{Soto13}. Then, we proceeded to a second selection using the CCD J-H vs K$_S$-[4.5], with the slope of the reddening band  $\frac{E_{J-H}}{E_{K_{S}-[4.5]}}= 2.07$ 
\citep{Chen13}. The third color selection was done using the CCD H-K$_S$ vs K$_S$-[4.5] with the slope of the reddening band $\frac{E_{J-K_{S}}}{E_{K_{S}-[4.5]}}= 0.98$ 
\citep{Chen13}. A total of 465 infrared excess sources passed all three color selections.
To then isolate Class I and Class II YSOs from these infrared excess sources, we used the Phase$\,$II color selection from \citet{Gutermuth09}, with the CCD [K$_S$-[3.6]]$_0$ vs [[3.6]-[4.5]]$_0$ as shown in Figure~\ref{fig:phaseII_classification}. We obtained a total of 24 Class I YSOs and a total of 557 Class II YSOs. 

Adding these sources then to what was found in Phase$\,$I, the final result of the infrared color selection method for YSOs with disks is a total of 66 Class I YSOs and a total of 733 Class II YSOs.

\section{X-ray and Infrared Associations}
\label{sec:X-ray and Infrared Associations}

\subsection{Infrared Counterparts of the X-Ray Sources}
\label{subsec:Infrared Counterparts of the X-Ray Sources}
\begin{figure}
\centering
\includegraphics[scale=0.31]{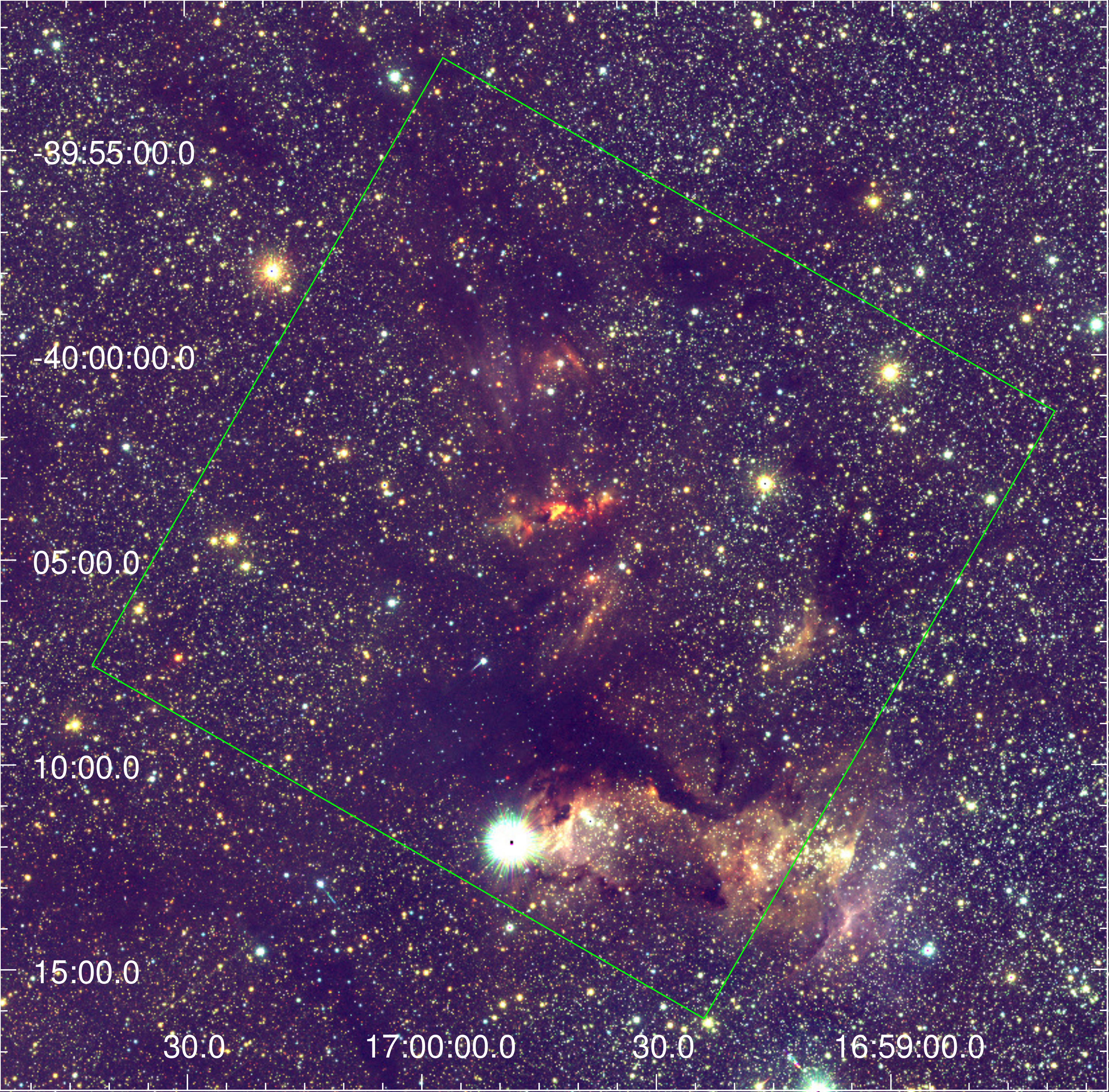}\includegraphics[scale=0.31]{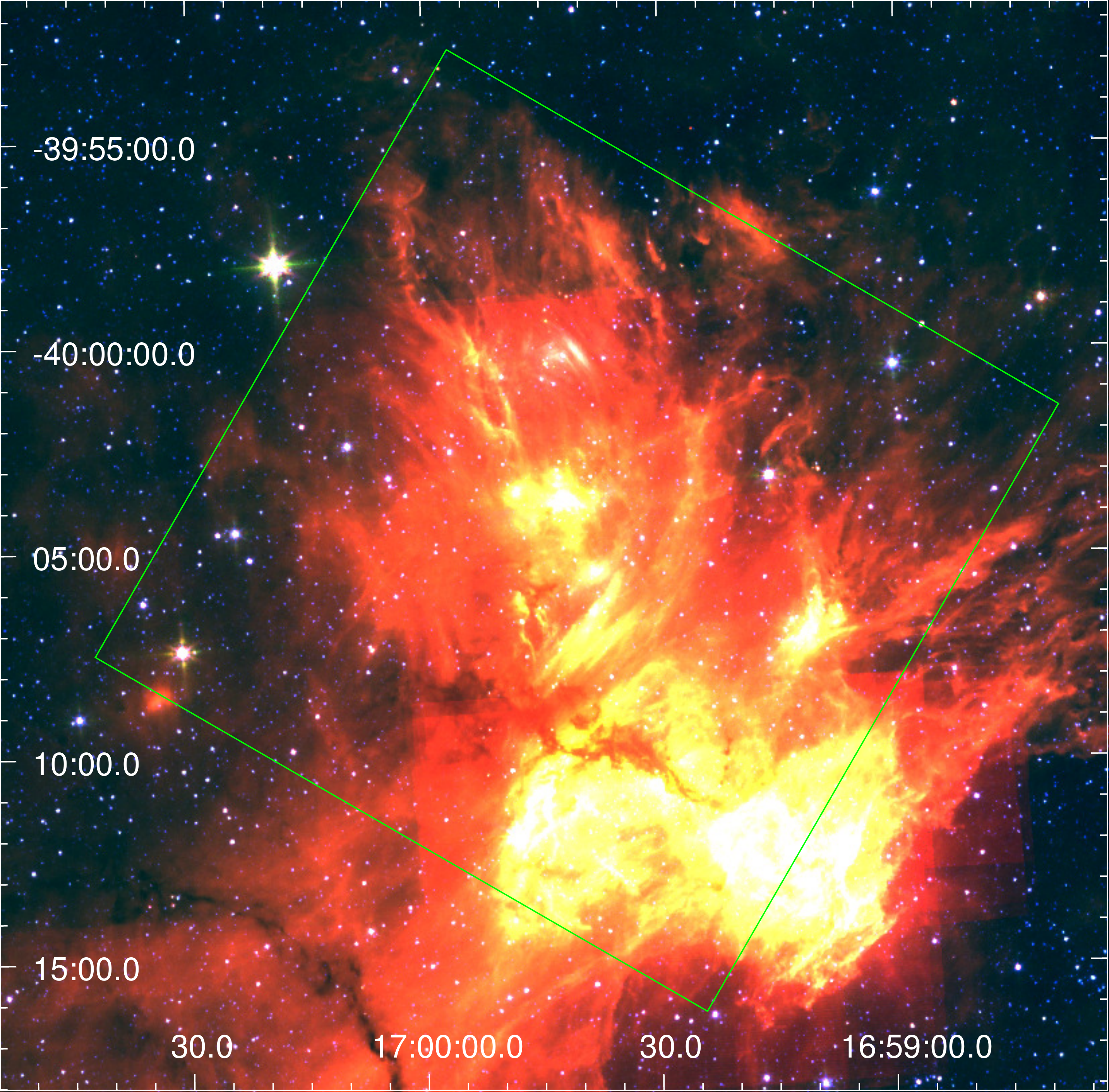}
\caption{\textit{Left}: VISTA/VVV survey three-color image with J-, H- and K$_S$-band composite of the IRAS 16562-3959 region. \textit{Right}: Spitzer/GLIMPSE survey three-color image with [3.6], [5.8] and [8.0] bands composite of the IRAS 16562-3959 region. The green square, for both images, indicates the position of the ACIS-I detector.}
\label{fig:VISTA_Spitzer}
\end{figure}

We searched for near- and mid- infrared counterparts of the 249 detected X-ray sources in the VISTA/VVV and Spitzer/GLIMPSE catalogs. In Figure~\ref{fig:VISTA_Spitzer}, we show three-color VISTA and Spitzer images of the IRAS 16562-3959 region, on which we have superposed the position of the ACIS array. The astrometric registration between these catalogs is better than 1$\arcsec$. Two different matching radii were used, because of the off-axis Chandra PSF degradation \citep{Getman05a}. We used 1$\arcsec$ for X-ray sources with off-axis position $\theta \leq$ 3$\arcmin$, and the matching radius was enlarged to 2$\arcsec$ for $\theta >$ 3$\arcmin$. When several counterparts of the same source were found, we chose the closest counterpart to the X-ray source. 
We found that 217 of the X-ray sources (87$\,\%$) have near-infrared counterparts. We selected the J, H and Ks bands, and kept for the analysis only the sources which have good detection in these 3 bands. We ended up with 165 counterparts, which corresponds to 66$\,\%$ of the X-ray sources. The sources and magnitudes are listed in Table~\ref{tab:IR_counterparts}.
Similarly, we found that 151 of the X-ray sources (61$\,\%$) have mid-infrared counterparts, which are listed with their magnitudes on Table~\ref{tab:IR_counterparts}. We kept for the analysis all the mid-infrared counterparts having at least one good detection in one of the bands [3.6], [4.5], [5.8] and [8.0] $\micron$. The analysis of the X-ray counterparts in terms of determination of population class and contamination is detailed in Section~\ref{sec:Infrared Analysis of the Region} together with the analysis of all infrared sources detected in the region. The results are presented in Table~\ref{tab:IR_counterparts}.

\subsection{Selection of Class III YSOs}
\label{subsec:Selection of Class III YSOs}

\begin{figure}
\centering
\includegraphics[scale=0.13]{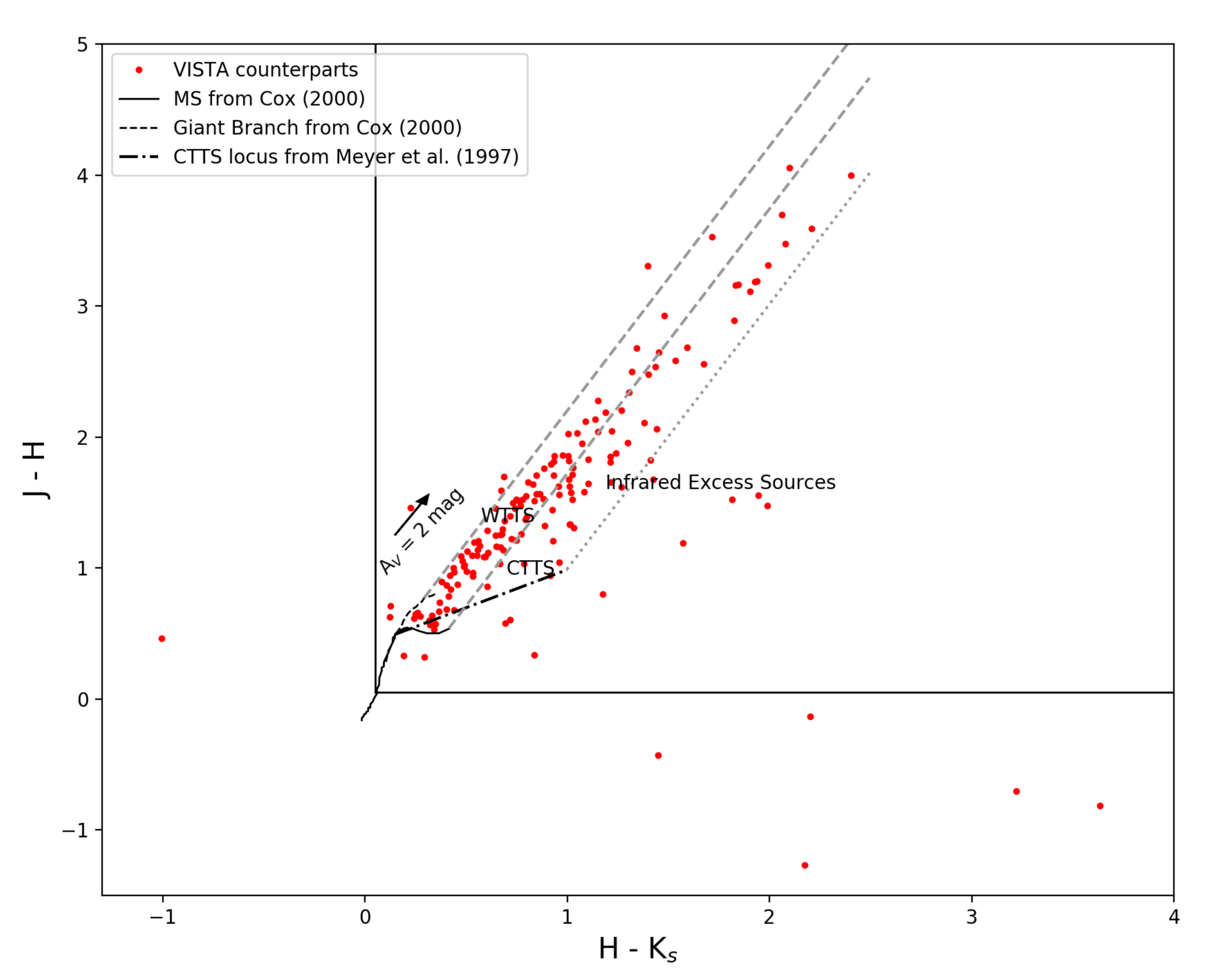}\includegraphics[scale=0.13]{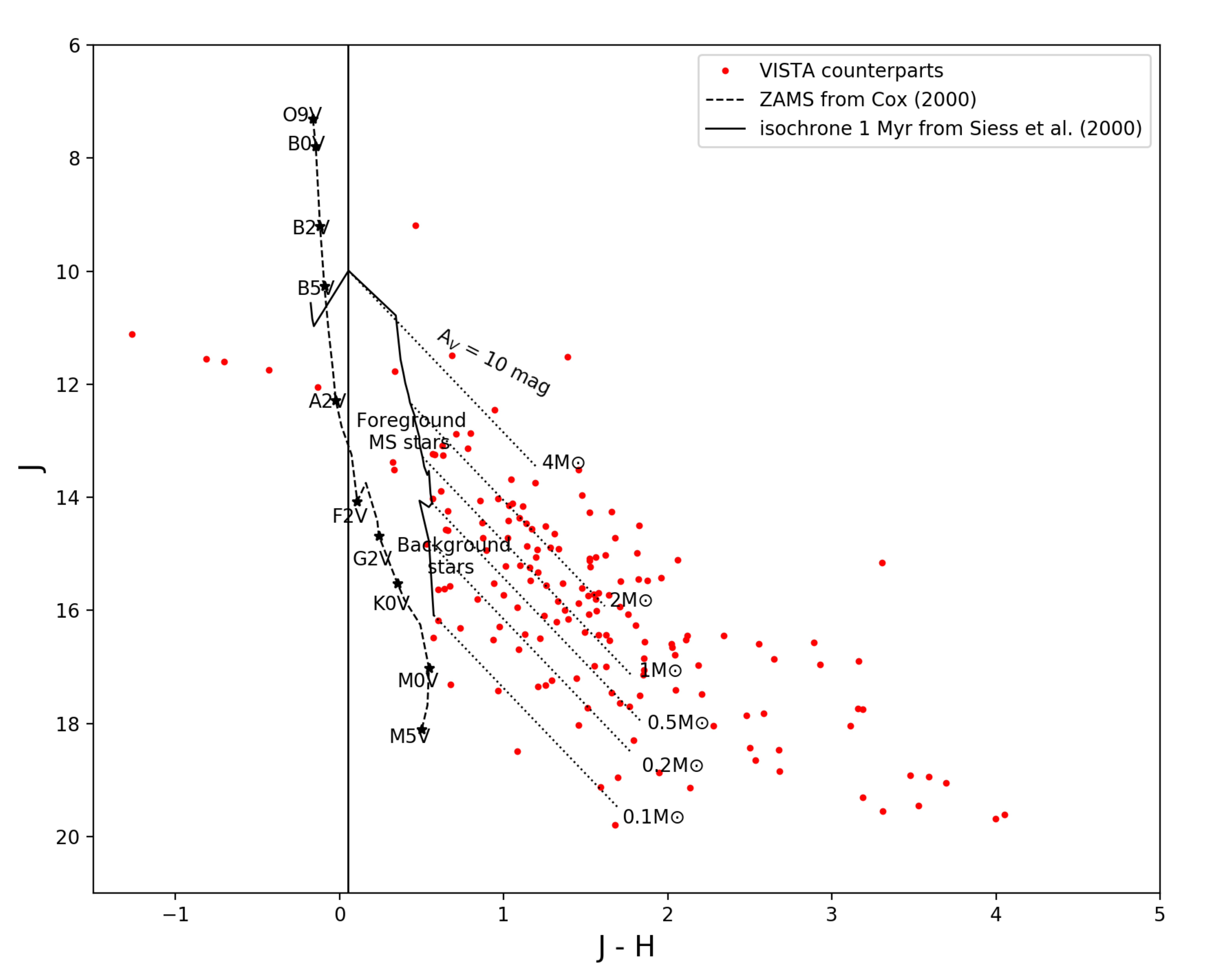}
\caption{\textit{Left:} CCD J-H vs H-K$_S$ of all X-ray sources having VISTA/VVV counterparts with high-quality photometry. The two straight black lines at (J-H) > 0.05 mag and (H-K$_S$) > 0.05\,mag \citep{Zeilder16} mark the limit for the infrared selection. The solid and dashed black line mark the intrinsic color position for main-sequence and giant stars from \citet{Cox00}. The gray dashed lines correspond to the reddening band for main-sequence stars from \citet{Rieke85}. The dash-dotted line show the locus for CTTS from \citet{Meyer97}, and the dotted line is the reddening band corresponding to the CTTS colors. \textit{Right:} CMD J vs J-H of the same stars as on the left panel. The straight black line at (J-H) > 0.05\,mag \citep{Zeilder16} mark the limit for the infrared selection. The dashed line indicates the location of the main-sequence stars at the distance to IRAS 16562-3959. The solid black line shows the 1\,Myr isochrone from \citet{Siess00}. Dotted lines show the reddening vectors with A$_V$ = 10\,mag for different stellar masses.}
\label{fig:VISTA_diagrams}
\end{figure}

We used our X-ray selected VISTA/VVV sources to identify Class III YSOs, 
by plotting the CCD J-H vs H-K$_S$ shown in Figure~\ref{fig:VISTA_diagrams}.
Sources defined as Class III YSOs or Weak-line T-Tauri Stars (WTTS) lack substantial
warm inner disk material, and are thus considered diskless, but they still could exhibit transitional, or debris disks.  They are located in the color space between the two gray dashed lines, where these sources do not exhibit any infrared excess, and thus can be explained by normal reddening. We found a total of 91 Class III YSOs. We also checked that the X-ray sources defined as Class II from the previous analysis are located in the color space defined by the reddened Classical T Tauri Star (CTTS, Class II) locus, and that the Class I sources are located on the right side of the dotted line, corresponding to the Class I color space. From this procedure we found 1 additional Class I YSO, and 31
additional Class II YSOs. We then proceeded to determine contamination using the color-magnitude diagram (CMD) J vs J-H shown in Figure~\ref{fig:VISTA_diagrams}, where the contamination from foreground and background stars is located to the left of the 1\,Myr isochrone.

\subsection{Unclassified X-Ray Sources} \label{sec:Unclassified X-Ray Sources}

\subsubsection{X-ray Sources Without Infrared Counterpart} \label{subsec:X-ray Sources Without Infrared Counterpart}
32 of our detected X-ray sources lack infrared counterparts. We consider two reasons for this. The first one is that those sources are located in regions of high extinction and cannot be detected in the infrared. The second one is that those sources could be extragalactic contamination, in particular AGNs. We consider sources that are located far from regions with high extinction or from the cluster, and that are very faint in X-ray, with count rate lower than 0.638 cts/ks as possible AGN contamination. We also have 3 sources (sources $\#$ 1, 226 and 248) that are located far from the cluster, but are very bright in X-ray. These sources are likely galactic contamination, such as X-ray binaries.

\subsubsection{X-ray Sources With Infrared Counterparts} \subsubsection{subsec:X-ray Sources With Infrared Counterparts}
We assume that the X-ray sources that have an incomplete set of infrared data are part of the cluster, and that the poor detection in infrared is due to high extinction. Some of these sources have good detections in the H and K$_S$ band only. A H-K$_S$ color study of our Class II and Class III sources indicates that the H-K$_S$ median for Class II sources is 1.27, and that for Class III sources it is 0.73. Therefore, we classify sources having H-K$_S$ < 1 as Class III, and sources having  H-K$_S$ > 1 as Class II.

\section{Subcluster Analysis}
\label{sec:Subcluster Analysis}

\subsection{Membership} \label{subsec:Membership}
\begin{figure}
\centering
\includegraphics[scale=0.165]{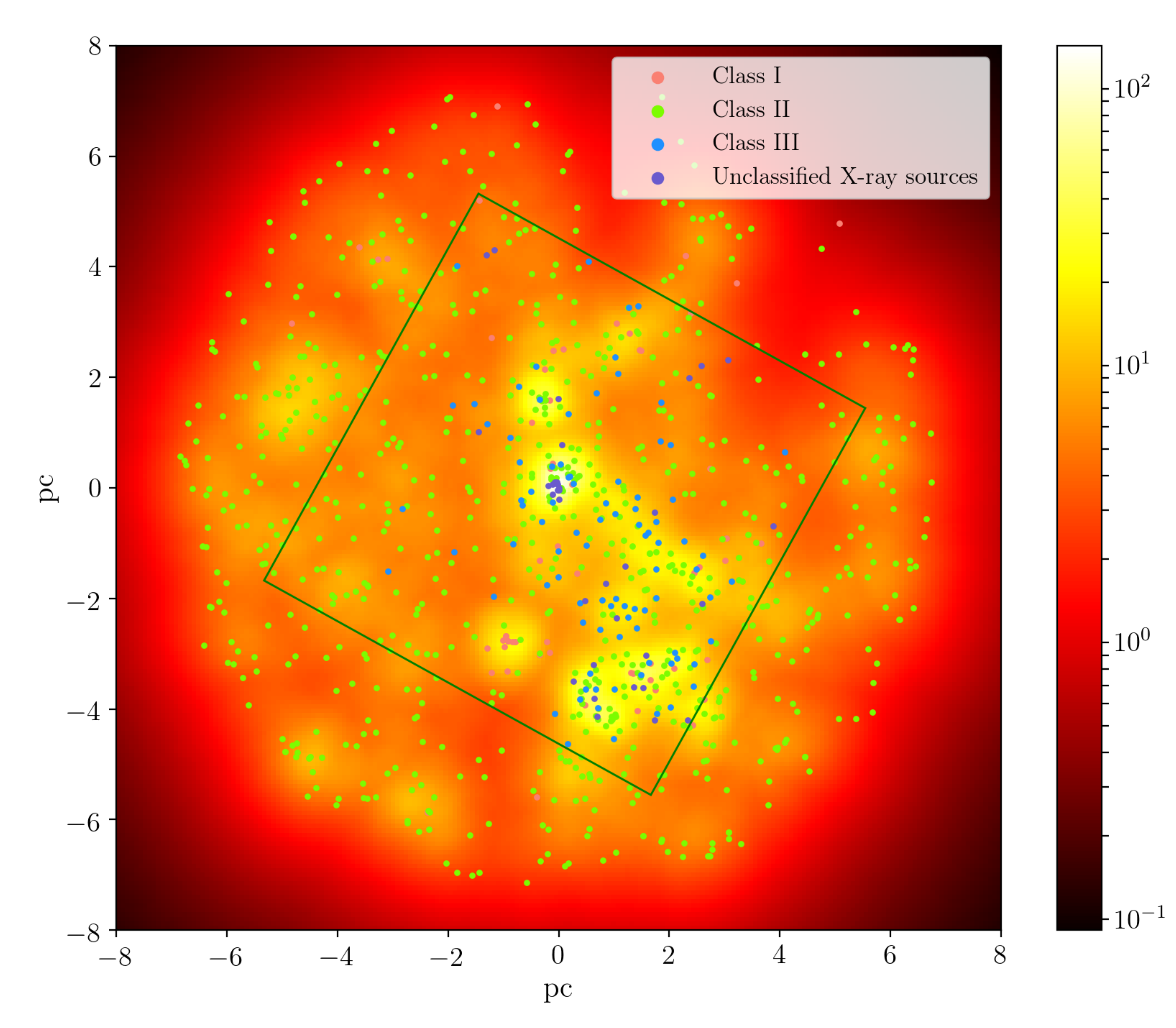}\includegraphics[scale=0.165]{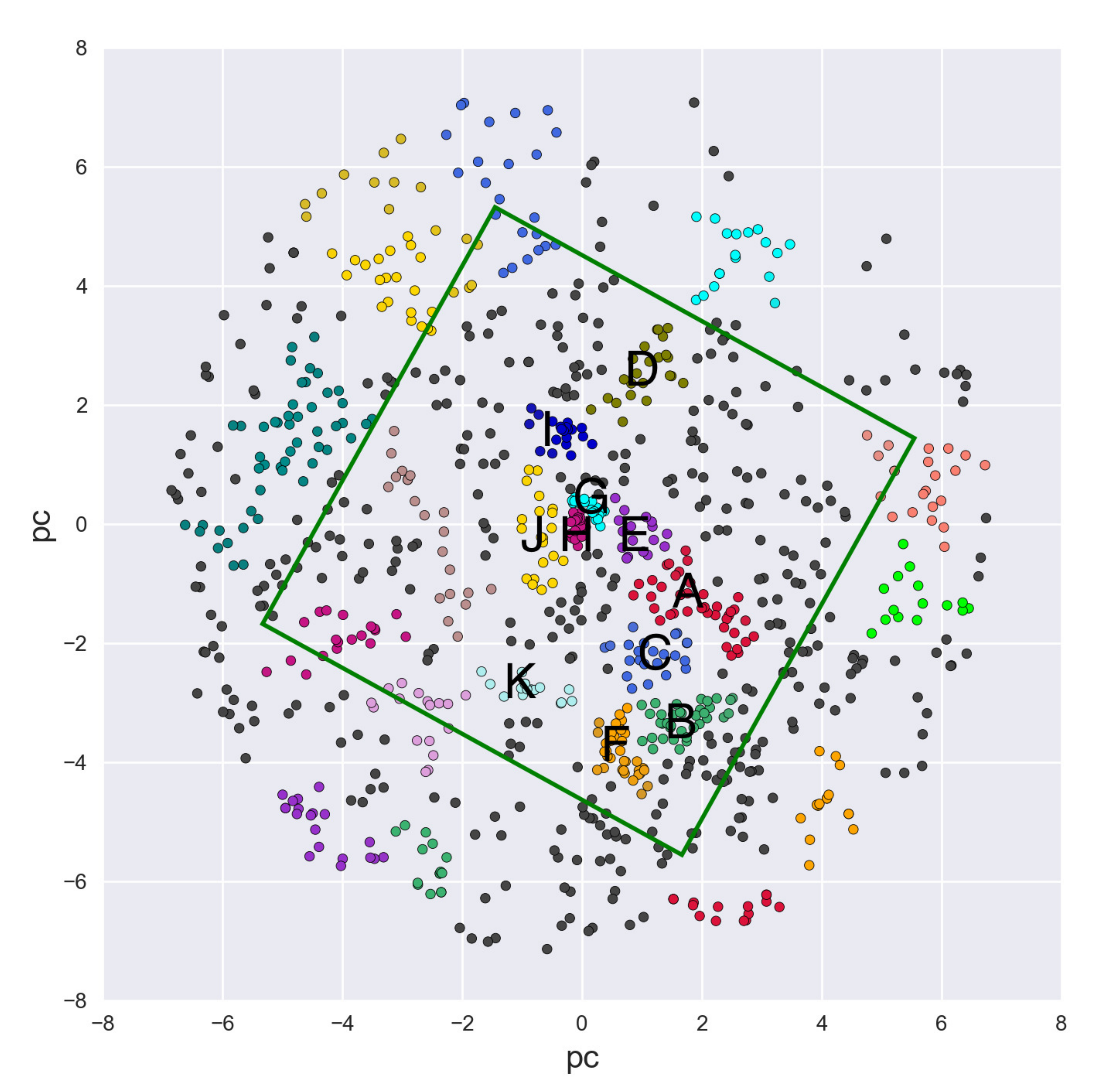}
\caption{\textit{Left:} Surface density plot for all YSOs considered to be part of the region, with each dot corresponding to a source. Red dots correspond to Class I YSOs, green dots to Class II, blue dots to Class III, and purple dots to unclassified X-ray sources. The units of the colors scale are number of sources per pc$^2$. The green square indicates the position of the ACIS-I detector. \textit{Right:} Cluster membership as determined with the HDBSCAN algorithm; each color corresponds to one cluster. The clusters selected for further analysis are denoted by letters.}
\label{fig:entire_field_clustering}
\end{figure}

Inspecting the images at different wavelength bands (Figures~\ref{fig:chandra} and \ref{fig:VISTA_Spitzer}), it is evident that the IRAS 16562-3959 region is not a simple centrally condensed cluster with the massive protostar at its center, but consist of several peak locations where stars have recently formed. In order to investigate the subcluster structure of the region, we have used the $k^{th}$-nearest neighbor density estimator (kNN) \citep{Casertano85}, with $k$=18, to obtain the YSO surface density. The result of this procedure is shown in Figure~\ref{fig:entire_field_clustering}, left panel. We have overlayed on this figure the location of all YSOs as determined in the previous sections. YSO surface density enhancements with the presence of Class I - III YSOs are clearly detected, together with a more uniform distribution of Class II objects. Note however, that our YSO selection method is not sensitive to Class III object outside the ACIS field, i.e. on mid/near-IR data alone. 

To further investigate the sub-cluster structure of IRAS 16562-3959, we used the hierarchical density based clustering  algorithm HDBSCAN \citep{McInnes17}, that uses the kNN algorithm, in association with the minimum spanning tree from Prim's algorithm, and hierarchical cluster analysis, to determine the clustering in a dataset (for more details see \textit{How HDBSCAN Works}\footnote{\url{http://hdbscan.readthedocs.io/en/latest/how_hdbscan_works.html}}). The resulting cluster selection from this algorithm is shown in 
Figure~\ref{fig:entire_field_clustering}, right panel. While there appears to be clustering throughout the region, for further analysis we will concentrate on the clusters that 
match the YSO density enhancements in Figure~\ref{fig:entire_field_clustering}, left panel, with a threshold of 13 sources pc$^{-2}$. We thus selected 11 dense clusters
which we have labeled with letters A - K. In Table~\ref{tab:cluster_structure_properties} we give the observed characteristics of each cluster including the name, central position
(the median position of all sources in the cluster), the cluster area using a convex hull algorithm which allows to find the smallest polygon that contains a group of discrete points, the radius calculated from the convex hull area, the surface density for each cluster, the total number of YSOs in each cluster and their classification. 

It is instructive to compare the subclusters in the IRAS\,16562-3959 region with similar objects studied previously. While IRAS\,16562-3959 clearly harbors a high-mass protostar 
which has already developed a luminosity approaching that of O-type ZAMS stars, the subclusters are generally smaller, and contain far less YSOs than similar high mass star forming young clusters, as for 
instance
the regions listed by \cite{Getman14}. Also, the subcluster structures in these latter regions show a large variation in both size and number of
stars compared to the subclusters we have defined above, which are much more uniform in appearance. Our subclusters appear much more similar to the clusters
studied by \citet{Gutermuth09},
in terms of physical size, and number of YSOs, and hence also in terms of YSO surface densities. Furthermore, the ratio of the number of Class II and Class I YSOs is similar.
Most of the brightest sources in the clusters studied by \citet{Gutermuth09} have IRAS luminosities below $10^4\,$L$_\odot$, indicating that no massive stars are present.   

It is further important to point out the high-mass protostar in IRAS 16562-3959 is located in subcluster H which has the highest observed YSO surface density. We determined a population of 20 YSOs, including the high-mass protostar, which were all detected in X-rays, and only 40\% of these have VISTA near-infrared counterparts. Subcluster H has the largest fraction of X-ray sources which were not detected at near- and mid-infrared wavelengths, and thus the most unclassified sources due to extinction. Clearly, the surface density associated with the high-mass protostar, as derived from the IR analysis, needs to be regarded as a lower limit.

\subsection{Subcluster X-ray Emission} \label{subsec:Subcluster X-ray Emission}

We have also measured the average X-ray emission per star in each subcluster. We added all counts in each subcluster region, and divided by the
sum of the exposure times to derive an average count rate spectrum. These data are shown in Figure~\ref{fig:clusters_spectra}. Subsequently, we used the same technique to fit 
the average X-ray spectrum for each subcluster with a thermal spectrum plus one absorption component as described in the section~\ref{subsec:spectroscopy}. The results of the fits are listed in Table~\ref{tab:clusters_X-ray_properties}.

\subsection{Age Estimation} \label{subsec:Age Estimation}

In order to estimate the age of each subcluster, we used the Age$_{JX}$ method described in \citet{Getman14}. Specifically we use equations
(1) and (2) of that paper to calculate the median age of each subcluster based on its median J-H value. In Table~7 we list for each subcluster
(column 1), the number of stars (column 2), the number of stars used in the median age calculation (column 3), the median J-H values (column 4),
and the median ages for each subcluster (column 5). The derived ages range bewteen $0.77\,$Myr (subcluster H) to $1.86\,$Myr (subcluster D).

We also calculated the disk fraction in each subcluster as follows: $Disk\ Fraction = \frac{N_{disk}}{(N_{disk} + N_{diskless})}$, with $N_{disk}$ the number of YSOs with disk (Class I and Class II), and $N_{diskless}$ the number of YSOs without disk (Class III). In Table~7, we list the number of stars used in the disk fraction calculation (column 6), the number of stars with and without disks (columns 7 and 8, respectively),
and in column 9 the disk fraction.

\section{Discussion}
\label{sec:Discussion}

In this study we have attempted to characterize the stellar population in IRAS$\,$16562-3959, in particular with emphasis on the stellar
environment of the high-mass protostar located in this region. Inspecting Figure~\ref{fig:VISTA_Spitzer}, it is clear that IRAS$\,$16562-3959 is located within a much
larger region of projected size $\geq10\,$pc, and the presence of many young stars in this area indicates that star formation has taken place throughout the region. In particular, there are several young clusters and H{\small II} regions toward the south-west of IRAS$\,$16562-3959, demonstrating that high-mass star formation has occurred. Given this morphology, with fully formed high-mass stars
located at 2.5 pc to the south-west of a high-mass protostar, one might ask whether triggered star formation has occurred in the region from west toward the east.
Considering, however, the ages derived for our subclusters, we see that there is no clear age gradient. For instance, while subcluster H, which contains the high-mass protostar, has in fact the youngest age ($0.77\,$Myr), the two oldest subclusters A and D (1.59 and 1.86$\,$Myr, respectively) are located on opposite sides of subcluster H. Thus, while the high-mass star formation to the south-west could have influenced and perhaps induced star formation in the region, a clear propagation direction is not evident.

The estimates of the ages for the identified subclusters in the region allow us to investigate a number of issues in the star formation history of the
region. First, the X-ray analysis presented in this and many similar works profits from the fact that young stars are far more X-ray luminous
than main sequence stars. The evolution toward lower X-ray luminosities with time is nicely documented in Figure~4 of \citet{Preibisch95}, which shows a decrease of median X-ray luminosity from about $10^{30.5}\,$erg$\,$s$^{-1}$ for young clusters to $10^{27.5}\,$erg$\,$s$^{-1}$ for main sequence field stars
within a time period of $4\,$Myr. With our data, we can check whether this trend is also present in the subclusters associated with the IRAS$\,$16562-3959 region. In Figure~\ref{fig:Lum_and_disk_freq_vs_age} a), we plot the average X-ray luminosity of each cluster versus age. While there is a large scatter, a trend of decreasing X-ray luminosity with age is
clearly visible. Second, many studies have attempted to use the disk fraction determined from NIR photometry as an age indicator
\citep[e.g.][]{Haisch01}. However, there appears to be a large spread of ages for disk lifetimes \citep[e.g.][]{Hernandez08}, for a variety of reasons, partly due to observational limitations, and partly due to environmental influences within the star forming clusters. Hence, considering our Figure~\ref{fig:Lum_and_disk_freq_vs_age} b), where we plot the disk fraction versus age, we do not detect a clear correlation. Third, our data allow us to check whether the clusters' stars are unbound, and that the cluster is hence expanding. In Figure~\ref{fig:Lum_and_disk_freq_vs_age} c), we plot the stellar surface density versus the age of the subcluster, and again observe a decreasing trend in the surface density with age. In particular, our youngest subcluster H, which contains the high mass protostar, has the highest surface density. 

\begin{figure}
\centering
\includegraphics[scale=0.29]{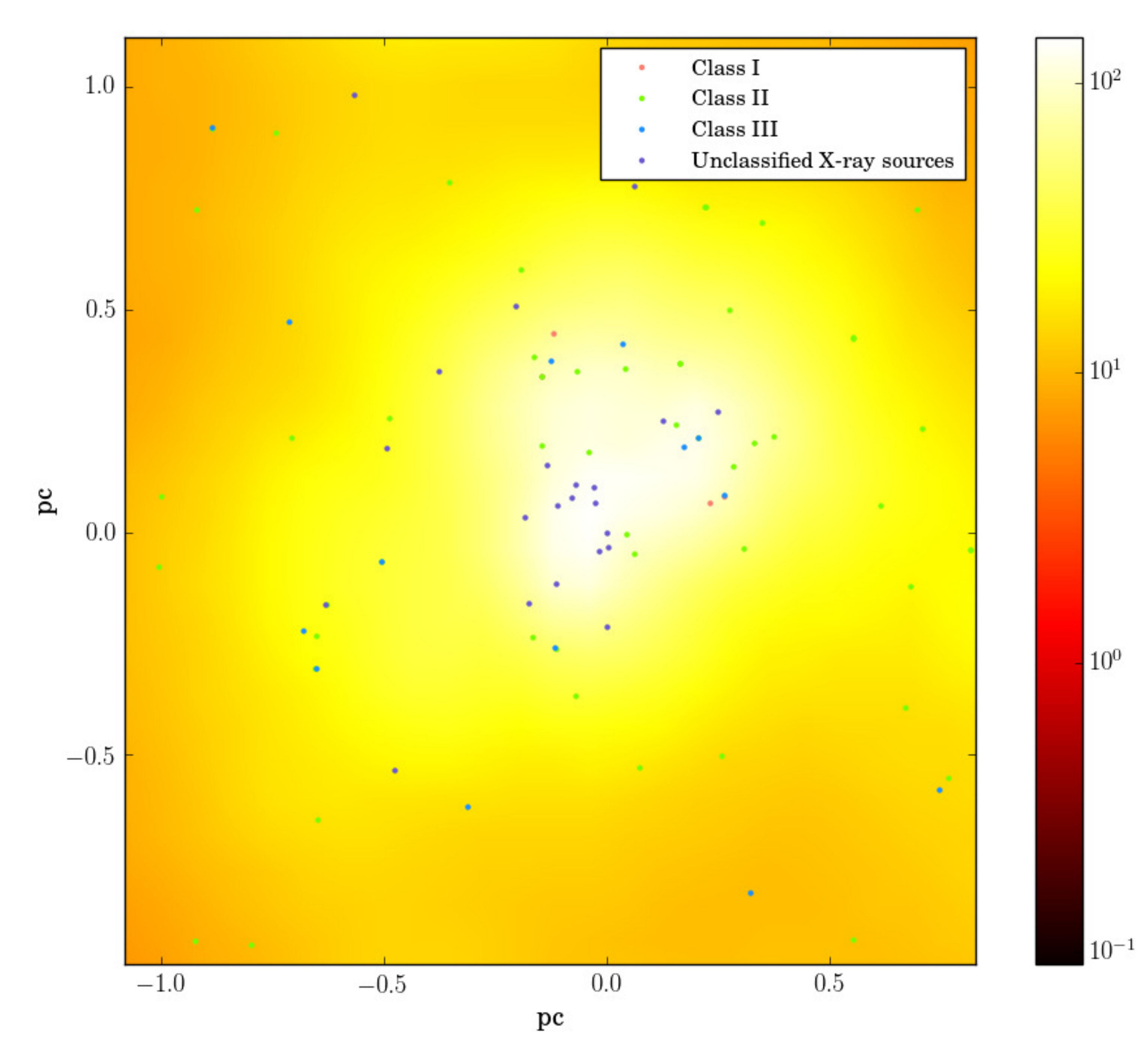}\includegraphics[scale=0.165]{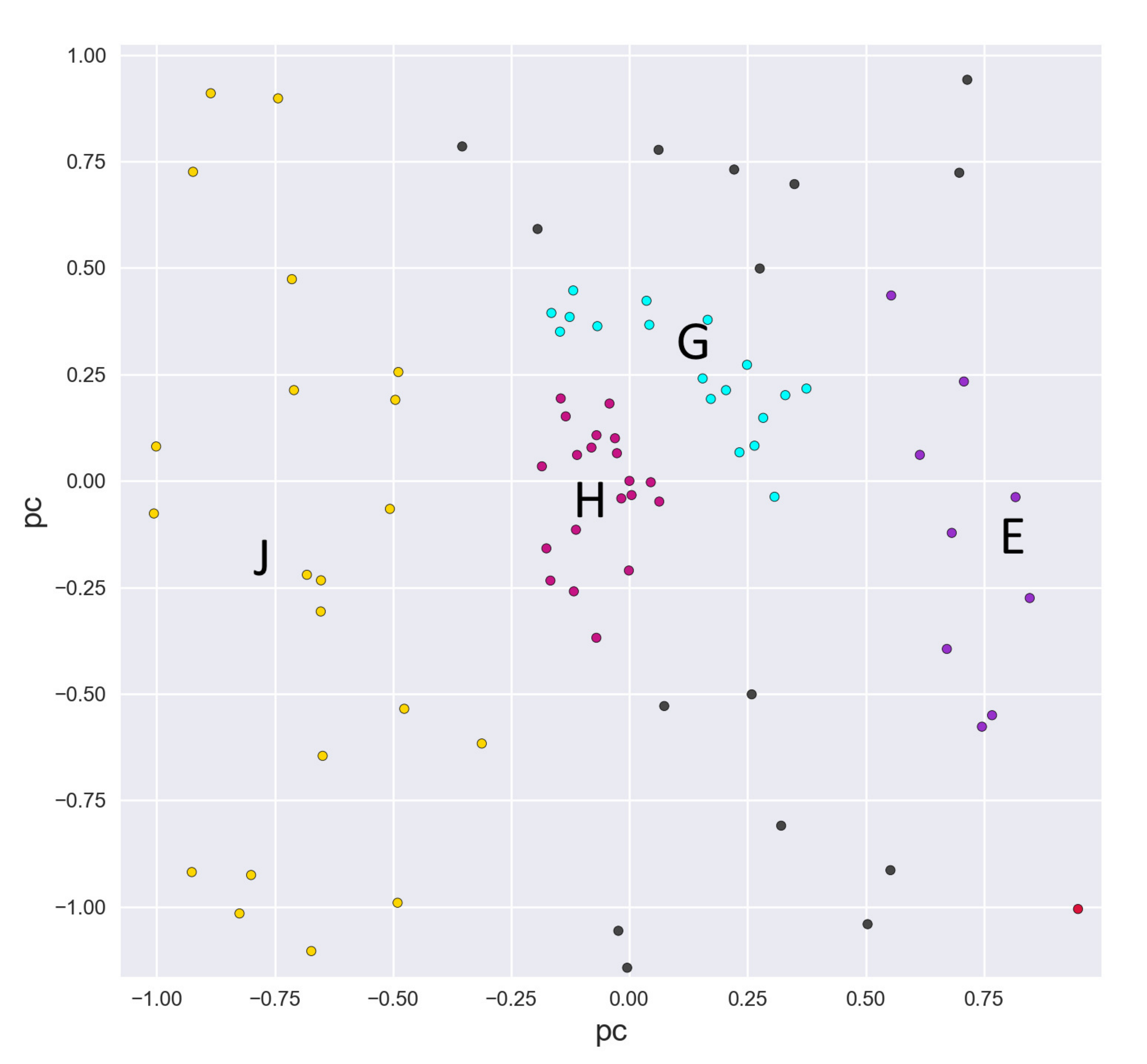}
\caption{\textit{Left:} Close up of the central cluster from the surface density plot of Figure~\ref{fig:entire_field_clustering}. \textit{Right:} Close up of the central cluster from the cluster membership plot of Figure~\ref{fig:entire_field_clustering}.}
\label{fig:central_cluster_clustering}
\end{figure}

\begin{sidewaysfigure}
\centering
\includegraphics[scale=0.2]{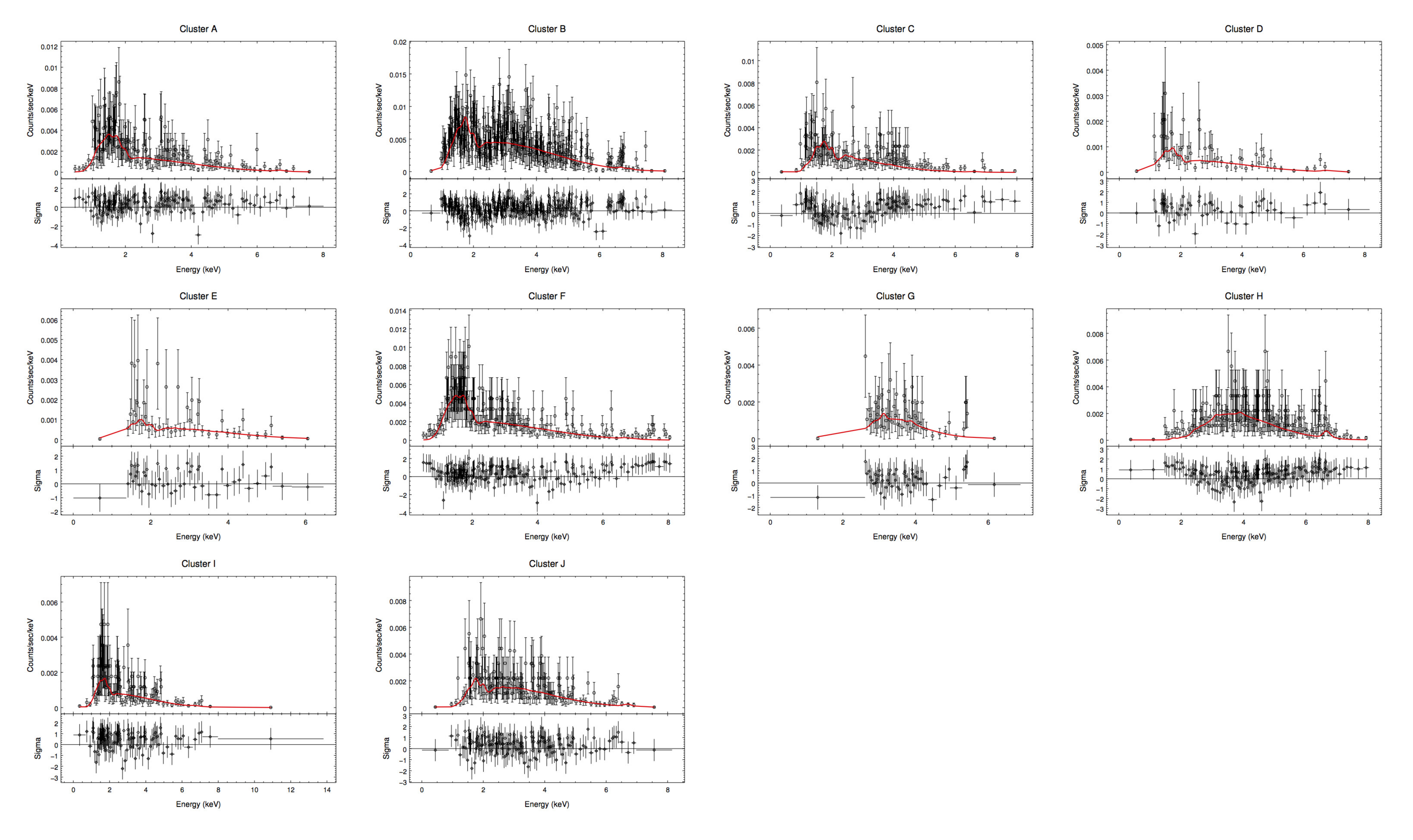}
\caption{Fitted average X-ray spectra for each subcluster.}
\label{fig:clusters_spectra}
\end{sidewaysfigure}

\begin{figure}
\centering
\includegraphics[scale=0.19]{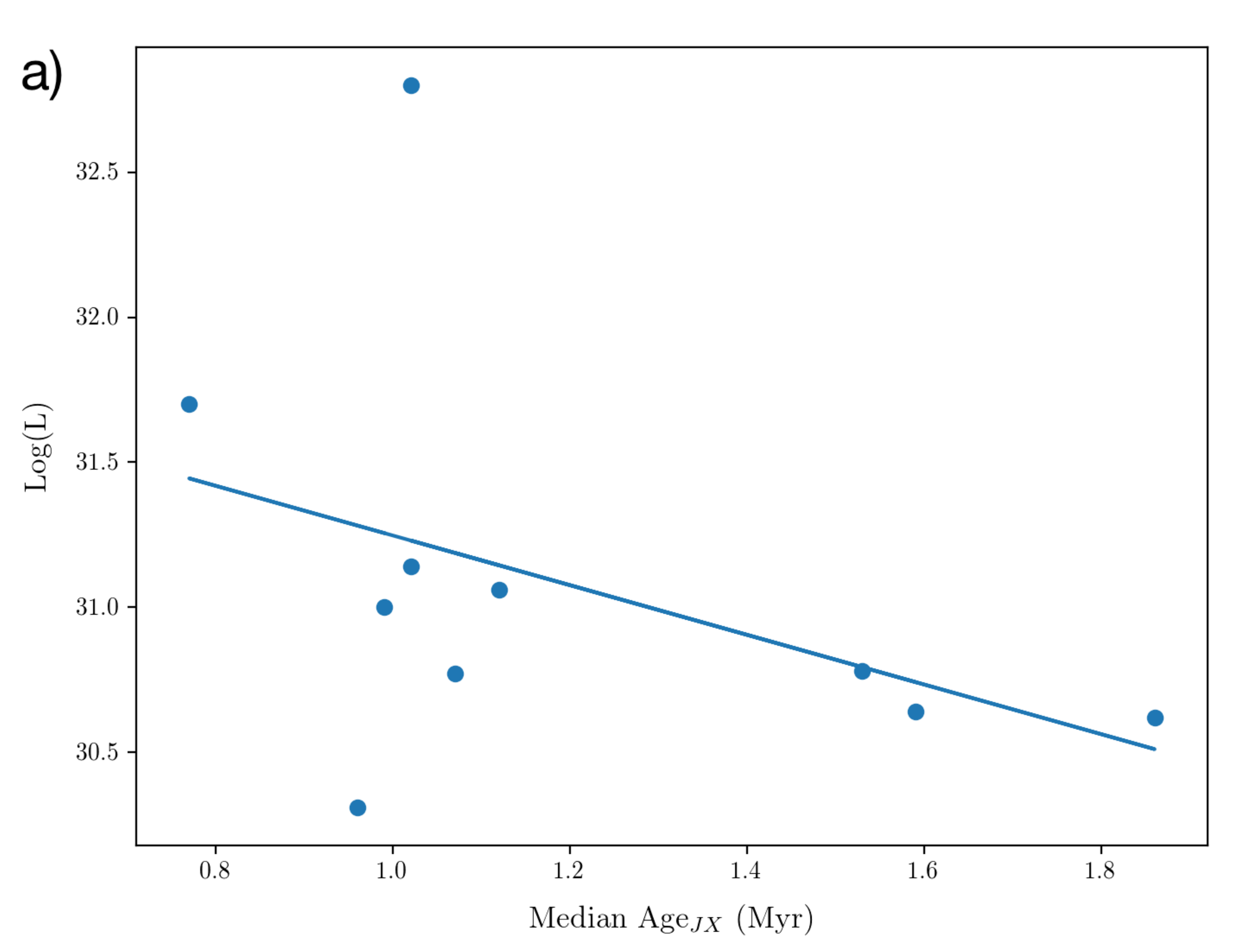}\\
\includegraphics[scale=0.19]{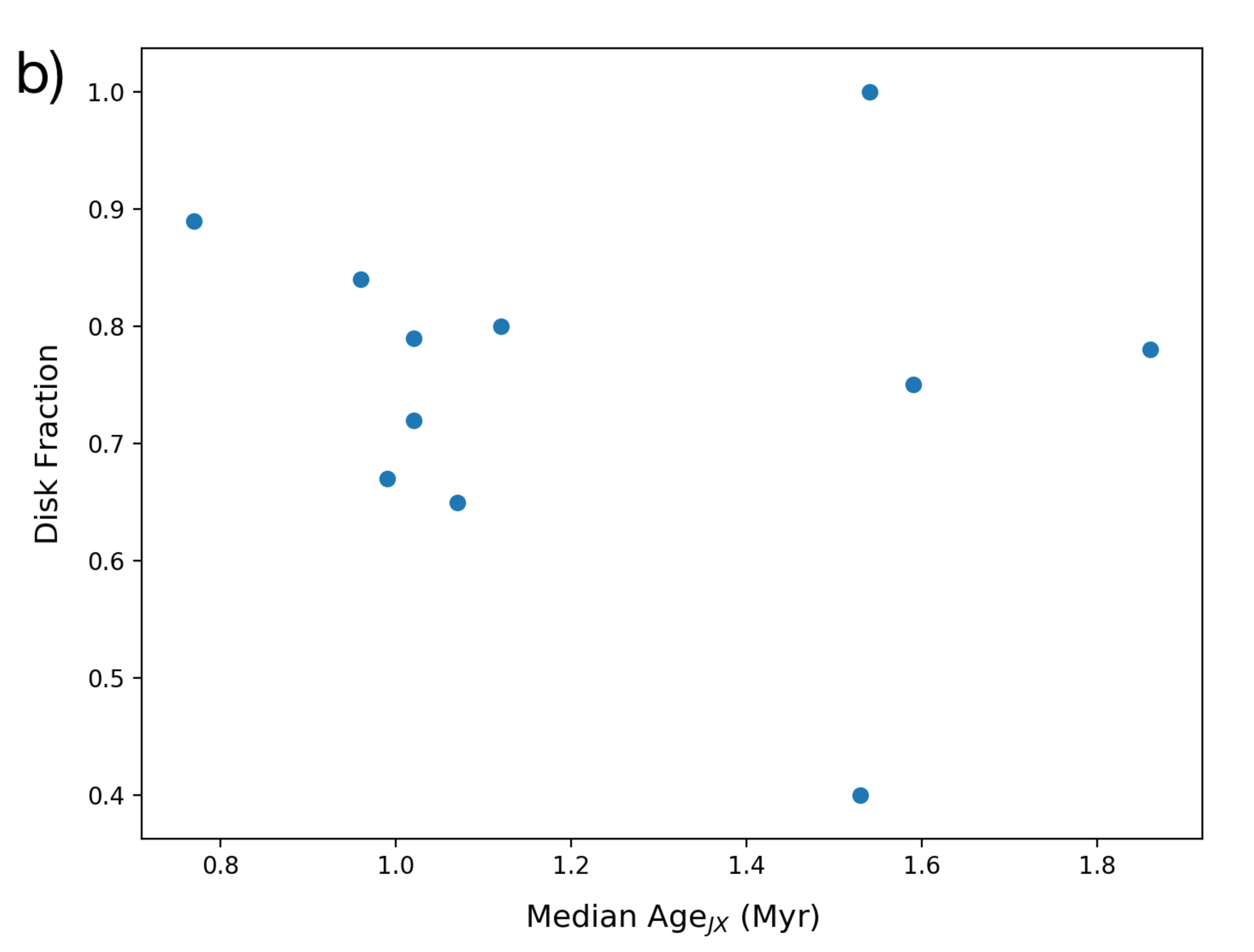}\\
\includegraphics[scale=0.19]{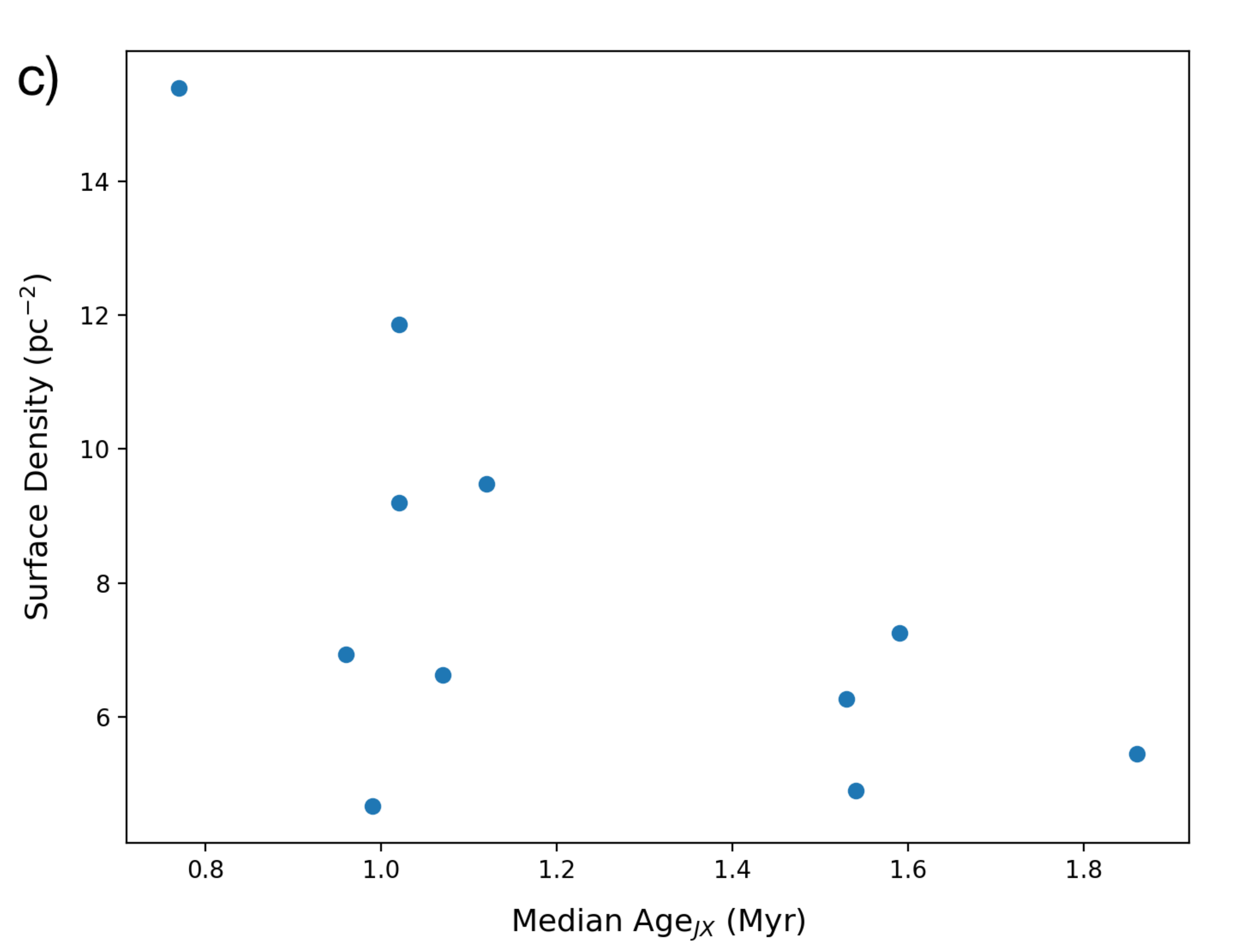}
\caption{\textit{a)} Plot of the log of the unabsorbed average luminosity of each subcluster vs. their median age, fitted with a linear regression. \textit{b)} Plot of the disk fraction calculated for each subcluster vs. their median age. \textit{c)} Plot of the surface density of each subcluster vs. their median age.}
\label{fig:Lum_and_disk_freq_vs_age}
\end{figure}

We will now turn to the immediate surroundings of the high-mass protostar G345.4938+01.4677. It is located in
subcluster H, which is the youngest and
most deeply embedded subcluster in this region. As pointed out above, it also has the highest surface density of young stellar objects, which
due to the large extinction toward this subcluster of N$_H \approx 10^{23}\,$cm$^{-2}$, is only a lower limit. We have detected an X-ray source
(source \# 161) coincident with the peak of radio continuum emission, which is presumably the position of the high-mass protostar. This
emission will be discussed in more detail in a future paper. The position of this object is coincident with the (0,0) position in Figures~\ref{fig:entire_field_clustering} and \ref{fig:central_cluster_clustering},
and from inspection of Figure~\ref{fig:central_cluster_clustering}, it is evident that the high-mass protostar is located near the center of subcluster H. Surrounding subcluster H are subclusters J, E, and G. We note that these three subclusters have ages of about $1\,$Myr, and all have lower surface densities than  subcluster H. We thus find that the high-mass protostar is embedded in a dense cluster, and surrounded with less dense and older distribution of stars, i.e. the stellar distribution can be described as a core-halo structure with an inside-out age gradient.

This result adds to recent findings of similar nature. For instance, a core-halo structure is evident in several clusters studied by \citet{Gutermuth09}, and recently \citet{Getman14b} reported core-halo age gradients in Orion and NGC$\,2024$. Another example is the study of IRAS 19343+2024 by \citet{Ojha10}: a high-mass protostellar cluster composed of at least four early B-type stars
(age $\sim$ 10$^4$ - 10$^5$ years) surrounded by a rich population of low-mass stars with ages 1 - 3 Myrs. While \citet{Getman14b} discussed a number of possible scenarios to explain the observed age gradient and core halo structure, the data presented in this work appear to favor the competitive accretion theory for high-mass star formation, where massive stars are expected to form later than the low mass stars in the cluster \citep{Tan14}.

Finally, it is worthwhile to consider our observational results with reference to a possible observational bias which could influence the judgment of whether the monolithic core accretion scenario of \citet{McKee03}, or the competitive accretion scenario of \citet{Bonnell01} is more applicable to the formation of high-mass stars. Candidates for high-mass protostars like the one studied in this paper are often found
by deep interferometric radio continuum surveys toward regions of high FIR luminosity, or massive molecular or dust cores. Because such observations are generally dynamic range limited, targets are often chosen from fields which are very radio quiet \citep[e.g.][]{Rosero16}, i.e. regions which are void of signs of previous high-mass star formation in the form of Ultra-compact or Hyper-compact H{\small II} regions.
In fact, radio observations of IRAS$\,$16562-3959 show no other detection of radio continuum sources, and hence possible high-mass (proto)stars at very low limits ($\mu$Jy) within $2\,$pc (\citet{Guzman16}, Montes et al. in prep). The early B-type protostar in IRAS$\,$20126+4104 is a similar 
example \citep[e.g.][]{Hofner07}. Hence, one might think that this selection method would bias against star formation theories which invoke stellar clusters, and bias toward high-mass stars which might form in isolation.
However, the present and other studies \citep[e.g.][]{Montes15} demonstrate that also in objects selected in this fashion, an older low-mass stellar cluster is present, indicating that the high-mass protostar may have been the last one to form in the cluster.

\section{Summary}
\label{sec:Summary}

We have performed a combined X-ray/infrared study of the young stellar population associated with the high-mass protostellar candidate in IRAS\,16562-3959. The main results of this study are:

\bigskip
1) We detected $249$ X-ray sources within the Chandra ACIS-I field-of-view. $70\%$ of them have a count rate lower than $0.638$\,cts/ks. The majority of the sources have a hard X-ray spectrum. We were able to obtain meaningful spectral fit results for 98 sources. The fitted N$_H$ values range from $2.1 \times 10^{21}$\,cm$^{-2}$ to $4.1 \times 10^{23}$\,cm$^{-2}$, and the corrected luminosities from $3.9 \times 10^{30}$\,erg$\,$s$^{-1}$ to $7.4 \times 10^{32}$\,erg$\,$s$^{-1}$, meaning that only relatively luminous X-ray sources were detected.

\bigskip
2) Phase I of the infrared analysis, using only the Spitzer/GLIMPSE data, resulted in the idenification of 218 YSOs, with 42 Class I and 176 Class II YSOs.
Phase II of the infrared analysis, combining Spitzer/GLIMPSE and VISTA/VVV data, yielded 581 YSOs, with 24 Class I and 557 Class II YSOs. Combining the results from Phase I and Phase II, it was found that a total of 799 YSOs are associated with the region, with 66 Class I, and 733 Class II YSOs.

\bigskip
3) We found that $87\%$ of the X-ray detections ($217$ sources) have near-infrared counterparts, but only $66\%$ ($165$ sources) have good detections in the J, H, and K$_s$ bands. We also found that $61\%$ of the X-ray detections ($151$ sources) have mid-infrared counterparts. The X-ray/infrared analysis resulted in the identification of 91 Class III YSOs.
Combining the results of the X-ray/infrared analysis, with the results
of the infrared analysis, we conclude that we have a total of $922$ YSOs associated with the region, with 67 Class I, $764$ Class II, and $91$ Class III YSOs.

\bigskip
4) We found that the region is composed of 11 subclusters. We determined the average X-ray luminosity, disk fraction, surface density, and age for each subcluster. We observed a decreasing trend in the average X-ray luminosity with age, no clear correlation between disk fraction and age, and a decreasing trend in the surface density with age. Also, triggered star formation is not evident in the region.

\bigskip
5) In the vicinity of the high-mass protostar, the stellar distribution has a core-halo structure. The subcluster in which the high-mass protostar was formed appears to be the youngest and has the highest surface density. The high-mass protostar (age $\sim$ 10$^5$ years), in its subcluster, is surrounded by older low-mass pre-main sequence stars (age $\sim$ 10$^6$ years), and it is located at the center. The results of the study are compatible with the competitive accretion model of \citet{Bonnell01}.

\acknowledgements

PH acknowledges support from SAO grants GO3-14005X, and GO5-16008X, as well as partial support from NSF grant AST-0908901 for this work. We thank the New Mexico Institute of Mining and Technology Research Office for their support.

\clearpage

\clearpage

\begin{deluxetable}{ccccccc}
	\tabletypesize{\scriptsize}
	    \tablewidth{0pc}
	\tablecaption{Log of Observations}
	\tablenum{1}
	\label{tab:log_chandra_obs}

	\tablehead{\colhead{Obs \#}&\colhead{Chandra Cycle}&\colhead{ID}&\colhead{Instrument}&\colhead{Exposure}&\colhead{Date}&\colhead{Roll Angle}}
\startdata
1	&	 14		&	ObsID 14537		&	ACIS-I		&	$5.02\,$ks		&	2013 Feb 2 	&	$95.45¡^\circ$ 	\\
2	&	 16		&	ObsID 16658		&	ACIS-I		&	$38.58\,$ks		&	2015 Jun 30 	&	$300.22¡^\circ$	\\
3	&	 16		&	ObsID 17691		&	ACIS-I		&	$40.07\,$ks		&	2015 Jul 3	 	&	$300.22¡^\circ$ 	\\
\enddata
\end{deluxetable}

\startlongtable
\begin{deluxetable}{ccccccccc}
	\tabletypesize{\scriptsize}
		    \tablewidth{0pc}
	\tablecaption{X-Ray Sources}
	\tablenum{2}
	\label{tab:X-ray_sources}

	\tablehead{\colhead{Source \#}&\colhead{R.A.}&\colhead{Dec.}&\colhead{Count Rate}&\colhead{F$_x$}&\colhead{HR}&\colhead{S.T. Var.}&\colhead{L. T.}&\colhead{Var.} \\
	\colhead{}&\colhead{(J2000)}&\colhead{(J2000)}&\colhead{(cts ks$^{-1}$)}&\colhead{10$^{-14}$}&\colhead{}&\colhead{Index Obs \#}&\colhead{Var.}&\colhead{} \\
	\colhead{}&\colhead{}&\colhead{}&\colhead{}&\colhead{(erg cm$^{-2}$ s$^{-1}$)}&\colhead{}&\colhead{1/2/3}&\colhead{Parameter}&\colhead{} \\
	\colhead{(1)}&\colhead{(2)}&\colhead{(3)}&\colhead{(4)}&\colhead{(5)}&\colhead{(6)}&\colhead{(7)}&\colhead{(8)}&\colhead{(9)}}
\startdata
1	&	16 58 52.73	&	-40 02 18.2	&	0.77	&	1.28	&	0.69	&	\nodata	/	6	/	0	&	1.49	&	STV		\\
2	&	16 58 55.13	&	-40 02 25.4	&	0.39	&	0.70	&	0.75	&	\nodata	/	1	/	\nodata	&	\nodata	&	\nodata		\\
3	&	16 58 56.35	&	-40 05 30.6	&	1.75	&	1.53	&	-0.31	&	\nodata	/	0	/	0	&	2.77	&	\nodata		\\
4	&	16 58 57.47	&	-40 05 07.6	&	0.98	&	2.57	&	0.65	&	\nodata	/	1	/	0	&	0.96	&	\nodata		\\
5	&	16 59 01.03	&	-40 01 50.1	&	0.44	&	0.56	&	-0.77	&	\nodata	/	1	/	\nodata	&	\nodata	&	\nodata		\\
6	&	16 59 01.87	&	-40 02 17.6	&	0.47	&	0.75	&	0.12	&	\nodata	/	\nodata	/	6	&	\nodata	&	STV		\\
7	&	16 59 04.58	&	-40 06 17.1	&	0.73	&	0.65	&	-0.17	&	\nodata	/	0	/	0	&	1.11	&	\nodata		\\
8	&	16 59 05.59	&	-40 06 24.0	&	1.49	&	1.59	&	-0.20	&	\nodata	/	6	/	0	&	0.05	&	STV		\\
\enddata
\end{deluxetable}

\clearpage

\startlongtable
\begin{deluxetable}{ccccccc}
	\tabletypesize{\scriptsize}
	    \tablewidth{0pc}
	\tablecaption{X-Ray Properties}
	\tablenum{3}
	\label{tab:X-ray_properties}

	\tablehead{\colhead{Source \#}&\colhead{N$_H$}&\colhead{kT}&\colhead{EM}&\colhead{F$_{x,c}$}&\colhead{log L$_{x,c}$}&\colhead{reduced $\chi^2$} \\
	\colhead{}&\colhead{10$^{22}$ (cm$^{-2}$)}&\colhead{(keV)}&\colhead{(cm$^{-3}$)}&\colhead{10$^{-14}$ (erg cm$^{-2}$ s$^{-1}$)}&\colhead{(erg s$^{-1}$)}&\colhead{} \\
	\colhead{(1)}&\colhead{(2)}&\colhead{(3)}&\colhead{(4)}&\colhead{(5)}&\colhead{(6)}&\colhead{(7)}}
\startdata
3	&	1.07	$	^	{	+	0.28	}	_	{	-0.23	}	$	&	0.78												&	10.58	$	^	{	+	3.52	}	_	{	-2.93	}	$	&	11.93	$	^	{	+	3.42	}	_	{	-3.57	}	$	&	31.62	&	0.60	\\
7	&	1.38	$	^	{	+	0.88	}	_	{	-0.58	}	$	&	1.23												&	3.09	$	^	{	+	2.21	}	_	{	-1.48	}	$	&	2.93	$	^	{	+	1.90	}	_	{	-1.72	}	$	&	31.01	&	0.64	\\
8	&	0.42	$	^	{	+	0.42	}	_	{	-0.31	}	$	&	2.51\tablenotemark{a}												&	2.31	$	^	{	+	0.90	}	_	{	-0.72	}	$	&	2.45	$	^	{	+	0.84	}	_	{	-0.77	}	$	&	30.93	&	0.73	\\
10	&	2.47	$	^	{	+	1.37	}	_	{	-0.89	}	$	&	2.52												&	3.18	$	^	{	+	1.48	}	_	{	-1.23	}	$	&	3.34	$	^	{	+	1.45	}	_	{	-1.46	}	$	&	31.06	&	0.74	\\
12	&	3.81	$	^	{	+	1.78	}	_	{	-1.35	}	$	&	1.09												&	9.08	$	^	{	+	6.97	}	_	{	-5.09	}	$	&	9.30	$	^	{	+	6.25	}	_	{	-5.94	}	$	&	31.51	&	0.43	\\
14	&	4.94	$	^	{	+	1.99	}	_	{	-1.37	}	$	&	1.88												&	14.68	$	^	{	+	6.36	}	_	{	-5.29	}	$	&	13.99	$	^	{	+	5.33	}	_	{	-5.63	}	$	&	31.68	&	0.51	\\
15	&	1.08	$	^	{	+	0.95	}	_	{	-0.54	}	$	&	0.50\tablenotemark{a}												&	7.96	$	^	{	+	10.67	}	_	{	-5.12	}	$	&	7.91	$	^	{	+	5.93	}	_	{	-6.08	}	$	&	31.44	&	1.08	\\
18	&	6.90	$	^	{	+	3.42	}	_	{	-2.02	}	$	&	1.87												&	10.16	$	^	{	+	5.31	}	_	{	-4.01	}	$	&	9.79	$	^	{	+	4.06	}	_	{	-4.55	}	$	&	31.53	&	0.95	\\
\enddata

\end{deluxetable}

\begin{longrotatetable}
\begin{deluxetable}{ccccccccc}
	\tabletypesize{\scriptsize}
	\tablecolumns{9}
	\tablecaption{Infrared Counterparts}
	\tablenum{4}
	\label{tab:IR_counterparts}

	\tablehead{\colhead{Source \#}&\colhead{J}&\colhead{H}&\colhead{K$_S$}&\colhead{[3.6]}&\colhead{[4.5]}&\colhead{[5.8]}&\colhead{[8.0]}&\colhead{Type} \\
	\colhead{(1)}&\colhead{(2)}&\colhead{(3)}&\colhead{(4)}&\colhead{(5)}&\colhead{(6)}&\colhead{(7)}&\colhead{(8)}&\colhead{(9)}}

\startdata
1	&	\nodata					&	\nodata					&	\nodata					&	\nodata					&	\nodata					&	\nodata					&	\nodata					&	Galactic contamination	\\
2	&	\nodata					&	17.58	$	\pm	$	0.09	&	16.99	$	\pm	$	0.09	&	\nodata					&	\nodata					&	\nodata					&	\nodata					&	Class III	\\
3	&	12.877	$	\pm	$	0.001	&	12.166	$	\pm	$	0.001	&	12.040	$	\pm	$	0.001	&	11.71	$	\pm	$	0.05	&	11.65	$	\pm	$	0.07	&	11.51	$	\pm	$	0.09	&	\nodata					&	Contamination	\\
4	&	\nodata					&	\nodata					&	17.61	$	\pm	$	0.16	&	\nodata					&	\nodata					&	\nodata					&	\nodata					&	\nodata	\\
5	&	14.244	$	\pm	$	0.003	&	13.584	$	\pm	$	0.003	&	13.324	$	\pm	$	0.004	&	13.03	$	\pm	$	0.07	&	13.32	$	\pm	$	0.11	&	\nodata					&	\nodata					&	Contamination	\\
6	&	\nodata					&	\nodata					&	\nodata					&	\nodata					&	\nodata					&	\nodata					&	\nodata					&	AGN	\\
7	&	\nodata					&	\nodata					&	\nodata					&	\nodata					&	\nodata					&	\nodata					&	\nodata					&	Class II	\\
8	&	11.774	$	\pm	$	0.001	&	11.436	$	\pm	$	0.001	&	10.601	$	\pm	$	0.001	&	10.56	$	\pm	$	0.09	&	10.40	$	\pm	$	0.10	&	10.16	$	\pm	$	0.06	&	9.98	$	\pm	$	0.06	&	Contamination	\\
\enddata
\end{deluxetable}
\end{longrotatetable}

\startlongtable
\begin{deluxetable}{ccccccccccccc}
	\tabletypesize{\scriptsize}
		    \tablewidth{0pc}
	\tablecaption{Subluster Structure Properties}
	\tablenum{5}
	\label{tab:cluster_structure_properties}

	\tablehead{\colhead{Subluster}&\colhead{R.A.}&\colhead{Dec.}&\colhead{A$_{hull}$}&\colhead{R}&\colhead{$\sigma_{hull}$}&\colhead{Total YSOs}&\colhead{Class I}&\colhead{Class II}&\colhead{Class III}&\colhead{Unclassified}&\colhead{Class II/}&\colhead{Class III/}\\
  \colhead{}&\colhead{(J2000)}&\colhead{(J2000)}&\colhead{(pc$^2$)}&\colhead{(pc)}&\colhead{(pc$^{-2}$)}&\colhead{}&\colhead{}&\colhead{}&\colhead{}&\colhead{}&\colhead{Class I}&\colhead{Class II} \\
	\colhead{(1)}&\colhead{(2)}&\colhead{(3)}&\colhead{(4)}&\colhead{(5)}&\colhead{(6)}&\colhead{(7)}&\colhead{(8)}&\colhead{(9)}&\colhead{(10)}&\colhead{(11)}&\colhead{(12)}&\colhead{(13)}}
\startdata
A	&	16 59 18.93	&	-40 06 35.5	&	5.38	&	1.31	&	7.25	&	39	&	2	&	25	&	9	&	3	&	12.50	&	0.36	\\
B	&	16 59 22.89	&	-40 10 15.6	&	3.91	&	1.12	&	9.20	&	36	&	6	&	20	&	7	&	3	&	3.33	&	0.35	\\
C	&	16 59 29.10	&	-40 08 09.1	&	3.67	&	1.08	&	6.27	&	23	&	0	&	8	&	12	&	3	&	$\infty$	&	1.50	\\
D	&	16 59 29.39 &	-39 58 41.2	&	4.22	&	1.16	&	5.45	&	23	&	4	&	14	&	5	&	0	&	3.50	&	0.36	\\
E	&	16 59 32.00	&	-40 04 03.7	&	2.57	&	0.90	&	6.62	&	17	&	0	&	11	&	6	&	0	&	$\infty$	&	0.55	\\
F	&	16 59 34.06	&	-40 11 38.6	&	3.59	&	1.07	&	9.48	&	34	&	1	&	23	&	6	&	4	&	23.00	&	0.26	\\
G	&	16 59 39.70	&	-40 03 12.5	&	1.52	&	0.70	&	11.86	&	18	&	3	&	10	&	5	&	0	&	3.33	&	0.50	\\
H	&	16 59 42.40	&	-40 03 43.9	&	1.30	&	0.64	&	15.40	&	20	&	0	&	8	&	1	&	10	&	$\infty$	&	0.13	\\
I	&	16 59 44.85	&	-40 00 30.9	&	2.89	&	0.96	&	6.93	&	20	&	3	&	13	&	3	&	1	&	4.33	&	0.23	\\
J	&	16 59 49.37	&	-40 04 10.4	&	4.51	&	1.20	&	4.66	&	21	&	1	&	13	&	7	&	0	&	13.00	&	0.54	\\
K	&	16 59 52.02	&	-40 09 20.5	&	3.06	&	0.99	&	4.90	&	15	&	10	&	5	&	0	&	0	&	0.50	&	$\infty$	\\
\enddata
\end{deluxetable}

\startlongtable
\begin{deluxetable}{ccccccc}
	\tabletypesize{\scriptsize}
	    \tablewidth{0pc}
	\tablecaption{Subcluster Average X-Ray Properties}
	\tablenum{6}
	\label{tab:clusters_X-ray_properties}

	\tablehead{\colhead{Subcluster}&\colhead{N$_H$}&\colhead{kT}&\colhead{EM}&\colhead{F$_{x,c}$}&\colhead{log L$_{x,c}$}&\colhead{reduced $\chi^2$} \\
	\colhead{}&\colhead{10$^{22}$ (cm$^{-2}$)}&\colhead{(keV)}&\colhead{(cm$^{-3}$)}&\colhead{10$^{-14}$ (erg cm$^{-2}$ s$^{-1}$)}&\colhead{(erg s$^{-1}$)}&\colhead{} \\
	\colhead{(1)}&\colhead{(2)}&\colhead{(3)}&\colhead{(4)}&\colhead{(5)}&\colhead{(6)}&\colhead{(7)}}
\startdata
A	&	0.65	$	^	{	+	0.17	}	_	{	-0.15	}	$	&	5.93	&	0.88	$	^	{	+	0.11	}	_	{	-0.11	}	$	&	1.27	$	^	{	+	0.16	}	_	{	-0.17	}	$	&	30.64	&	0.77	\\
B	&	1.38	$	^	{	+	0.18	}	_	{	-0.18	}	$	&	10.85	&	2.50	$	^	{	+	0.20	}	_	{	-0.20	}	$	&	3.95	$	^	{	+	0.32	}	_	{	-0.34	}	$	&	31.14	&	0.90	\\
C	&	2.13	$	^	{	+	0.50	}	_	{	-0.40	}	$	&	2.00	&	1.80	$	^	{	+	0.40	}	_	{	-0.34	}	$	&	1.73	$	^	{	+	0.35	}	_	{	-0.34	}	$	&	30.78	&	0.83	\\
D	&	1.29	$	^	{	+	0.52	}	_	{	-0.40	}	$	&	5.50	&	0.84	$	^	{	+	0.21	}	_	{	-0.19	}	$	&	1.20	$	^	{	+	0.29	}	_	{	-0.28	}	$	&	30.62	&	0.77	\\
E	&	2.05	$	^	{	+	0.70	}	_	{	-0.51	}	$	&	3.16	&	1.47	$	^	{	+	0.39	}	_	{	-0.35	}	$	&	1.70	$	^	{	+	0.48	}	_	{	-0.44	}	$	&	30.77	&	0.67	\\
F	&	0.71	$	^	{	+	0.13	}	_	{	-0.11	}	$	&	8.62	&	2.12	$	^	{	+	0.19	}	_	{	-0.19	}	$	&	3.28	$	^	{	+	0.32	}	_	{	-0.31	}	$	&	31.06	&	0.91	\\
G	&	17.40	$	^	{	+	4.10	}	_	{	-3.25	}	$	&	0.69	&	164.58	$	^	{	+	87.93	}	_	{	-58.56	}	$	&	181.57	$	^	{	+	82.11	}	_	{	-82.07	}	$	&	32.80	&	0.63	\\
H	&	11.88	$	^	{	+	2.39	}	_	{	-1.96	}	$	&	2.00	&	15.11	$	^	{	+	3.57	}	_	{	-2.93	}	$	&	14.50	$	^	{	+	3.32	}	_	{	-3.15	}	$	&	31.70	&	0.83	\\
\enddata
\end{deluxetable}

\clearpage
\startlongtable
\begin{deluxetable}{ccccccccc}
	\tabletypesize{\scriptsize}
		    \tablewidth{0pc}
	\tablecaption{Age and Disk Fraction Estimates for Each Subcluster}
	\tablenum{7}
	\label{tab:age_disk_fraction}

	\tablehead{\colhead{Subcluster}&\colhead{N$_{tot}$}&\colhead{N$_{Age}$}&\colhead{Median}&\colhead{Median}&\colhead{N$_{disk frac}$}&\colhead{N$_{disk}$}&\colhead{N$_{diskless}$}&\colhead{Disk}\\
	\colhead{}&\colhead{}&\colhead{}&\colhead{(J-H)}&\colhead{Age (Myr)}&\colhead{}&\colhead{}&\colhead{}&\colhead{Fraction} \\
	\colhead{(1)}&\colhead{(2)}&\colhead{(3)}&\colhead{(4)}&\colhead{(5)}&\colhead{(6)}&\colhead{(7)}&\colhead{(8)}&\colhead{(9)}}
\startdata
A	&	39	&	34	&	1.41	&	1.59	$	_	{	-	0.64	}	^	{	+	1.41	}	$	&	36	&	27	&	9	&	0.75	\\
B	&	36	&	24	&	1.87	&	1.02	$	_	{	-	0.35	}	^	{	+	1.08	}	$	&	33	&	26	&	7	&	0.79	\\
C	&	23	&	19	&	1.44	&	1.53	$	_	{	-	0.63	}	^	{	+	1.52	}	$	&	20	&	8	&	12	&	0.4	\\
D	&	23	&	9	&	1.25	&	1.86	$	_	{	-	0.86	}	^	{	+	1.69	}	$	&	23	&	18	&	5	&	0.78	\\
E	&	17	&	15	&	1.76	&	1.07	$	_	{	-	0.37	}	^	{	+	1.13	}	$	&	17	&	11	&	6	&	0.65	\\
F	&	34	&	28	&	1.68	&	1.12	$	_	{	-	0.37	}	^	{	+	1.13	}	$	&	30	&	24	&	6	&	0.8	\\
G	&	18	&	15	&	1.88	&	1.02	$	_	{	-	0.35	}	^	{	+	1.03	}	$	&	18	&	13	&	5	&	0.72	\\
H	&	20	&	8	&	2.49	&	0.77	$	_	{	-	0.27	}	^	{	+	0.83	}	$	&	9	&	8	&	1	&	0.89	\\
\enddata
\tablecomments{(1) Cluster designation. (2) Total number of YSOs in each cluster. (3) Number of YSOs having near-infrared counterparts used for the age calculation. (4) median color (J-H) for all stars having near-infrared counterparts in each cluster. (5) Median age calculated for all stars having near-infrared counterparts in each cluster. The uncertainties are the 25$\%$ and 75$\%$ quartiles of
the Age$_{JX}$ versus J-H distribution from Getman et al. (2014). (6) Number of YSOs used for the disk fraction calculation. (7) Number of YSOs with disk (Class I and Class II). (8) Number of YSOs without disk (Class III). (9) Disk fraction calculated for each cluster; see text for more details.}
\end{deluxetable}

\end{document}